\documentclass[preprint2]{aastex6}

\begin{document}
\title{Signs of Early-Stage Disk Growth Revealed with ALMA}

\author{Hsi-Wei Yen\altaffilmark{1,2}, Patrick M. Koch\altaffilmark{1}, Shigehisa Takakuwa\altaffilmark{3,1}, Ruben Krasnopolsky\altaffilmark{1}, Nagayoshi Ohashi\altaffilmark{4,1}, Yusuke Aso\altaffilmark{5}}
\altaffiltext{1}{Academia Sinica Institute of Astronomy and Astrophysics, P.O. Box 23-141, Taipei 10617, Taiwan} 
\altaffiltext{2}{Current address: European Southern Observatory (ESO), Karl-Schwarzschild-Str. 2, D-85748 Garching, Germany; hyen@eso.org}
\altaffiltext{3}{Department of Physics and Astronomy, Graduate School of Science and Engineering, Kagoshima University, 1-21-35 Korimoto, Kagoshima, Kagoshima 890-0065, Japan}
\altaffiltext{4}{Subaru Telescope, National Astronomical Observatory of Japan, 650 North A'ohoku Place, Hilo, HI 96720, USA}
\altaffiltext{5}{Department of Astronomy, Graduate School of Science, The University of Tokyo, 731 Hongo, Bunkyo-ku, Tokyo 113-0033, Japan}

\begin{abstract}
We present ALMA 1.3 mm continuum, $^{12}$CO, C$^{18}$O, and SO data for the Class 0 protostars, Lupus 3 MMS, IRAS 15398$-$3559, and IRAS 16253$-$2429 at resolutions of $\sim$100 AU.  By measuring a rotational profile in C$^{18}$O, a 100 AU Keplerian disk around a 0.3 $M_\sun$ protostar is observed in Lupus 3 MMS. No 100 AU Keplerian disks are observed in IRAS 15398$-$3559 and IRAS 16253$-$2429. Nevertheless, embedded compact ($<$30 AU) continuum components are detected. The C$^{18}$O emission in IRAS 15398$-$3559 shows signatures of infall with a constant angular momentum.  IRAS 16253$-$2429 exhibits signatures of infall and rotation, but its rotational profile is unresolved.  By fitting the C$^{18}$O data with our kinematic models, the protostellar masses and the disk radii are inferred to be 0.01 $M_\sun$ and 20 AU in IRAS 15398$-$3559, and 0.03 $M_\sun$ and 6 AU in IRAS 16253$-$2429. By comparing the specific angular momentum profiles from 10,000 to 100 AU in 8 Class 0 and I protostars, we find that the evolution of envelope rotation can be described with conventional inside-out collapse models. In comparison with a sample of 18 protostars with known disk radii, our results reveal signs of disk growth, with the disk radius increasing as ${M_*}^{0.8\pm0.14}$ or $t^{1.09\pm0.37}$ in the Class 0 stage, where $M_*$ is the protostellar mass and $t$ is the age. The disk growth rate slows down in the Class I stage.  Besides, we find a hint that the mass accretion rate declines as $t^{-0.26\pm0.04}$ from the Class 0 to I stages.
\end{abstract}

\section{Introduction}
Keplerian disks with radii of hundreds of AU are often observed around young stellar objects, such as T Tauri or Herbig Ae/Be stars, which are considered sites of planet formation (Williams \& Cieza 2011). 
Similar disks are observed in molecular lines around several sources at earlier evolutionary stages, Class 0 and I protostars. 
From the gas kinematics traced by molecular lines,
the Keplerian disks detected around Class I protostars have radii ranging from 50 AU to larger than 700 AU, and their protostellar masses range from 0.5 $M_\sun$ to 2.5 $M_\sun$ (Lommen et al.~2008; Takakuwa et al.~2012; Brinch \& J{\o}rgensen et al.~2013; Chou et al.~2014; Harsono et al.~2014; Lindberg et al.~2014; Yen et al.~2014; Aso et al.~2015; Lee et al.~2016). 
The radii of Keplerian disks around Class 0 protostars are 50 AU to 150 AU with protostellar masses of 0.2--0.3 $M_\sun$ (Tobin et al.~2012a; Murillo et al.~2013; Ohashi et al.~2014; Lee et al.~2014). 
In addition, recent JVLA surveys in 8 mm continuum also found several candidate disks with radii of tens of AU around Class 0 and I protostars by analyzing the continuum visibility amplitude profiles (Segura-Cox et al.~2016). 
In contrast to this,  
there is a group of Class 0 protostars that exhibit one order of magnitude slower envelope rotations than other Class 0 protostars on a 1000 AU scale, 
suggesting that their radii of Keplerian disks are likely less than 10 AU (Brinch et al.~2009; Yen et al.~2010, 2013, 2015a; Maret et al.~2014).
These results show that Class 0 and I protostars clearly exhibit a wide range of disk sizes. 
It is, therefore, still unclear as to when and how Keplerian disks form and grow to those larger-than-100 AU disks seen around young stellar objects.

Keplerian disks are expected to form when collapsing material rotates fast enough to become centrifugally supported (e.g., Terebey et al.~1984).
Conventionally, Keplerian disks are expected to grow in size as the collapse proceeds because more angular momentum is transferred to the disk-forming region due to the conservation of angular momentum. 
In the inside-out collapse model of non-magnetized rigid rotating dense cores, 
the radius of a Keplerian disk grows $\propto {M_{\rm sd}}^3$, where $M_{\rm sd}$ is the total mass of the star+disk systems (Terebey et al.~1984).
In the model considering dense core formation in magnetized rigid rotating clouds, 
the initial radial profile of angular velocity ($\omega$) approaches $\omega \propto r^{-1}$, as the dense cores become supercritical (Basu \& Mouschovias 1994). 
The collapse of these dense cores then proceeds conserving angular momentum, 
and the radius of the Keplerian disk grows $\propto M_{\rm sd}$ (Basu 1998). 
On the other hand, 
ideal magneto-hydrodynamical (MHD) simulations of the collapse of dense cores, that initially exhibit aligned rotational and magnetic field axes, show that the magnetic field efficiently removes angular momentum from the collapsing material via magnetic braking, 
and thus,  no Keplerian disks larger than 10 AU can form (e.g., Allen et al.~2003; Mellon \& Li 2008).
That contradicts, at least, some observations that show an increasing number of Keplerian disks with radii larger than tens of AU.
To reduce the efficiency of magnetic braking, 
simulations additionally consider non-ideal MHD effects, dissipation of protostellar envelopes, initially misaligned rotational axis and magnetic field, more realistic treatment of ionization degrees, or turbulence (e.g., Hennebelle \& Ciardi 2009; Li et al.~2011, 2013; Machida \& Matsumoto 2011; Machida et al.~2011,2014; Dapp et al.~2012; Santos-Lima et al.~2012; Seifried et al.~2013, 2013; Joos et al.~2012, 2013; Padovani et al.~2013, 2014; Tomida et al.~2015; Zhao et al.~2016). 
Several of these simulations show that Keplerian disks with tens of AU in size can form in magnetized dense cores.  
However, it is still unknown which mechanisms play a more important or possibly dominating role in the process of disk formation and growth. 
Observations revealing the evolution of disk sizes are essential to shed light on which mechanisms are dominant. 

The current observational results suggest that (1) there is likely an evolutionary trend from slow, to fast, and to Keplerian rotation on a scale of hundreds of AU around Class 0 and I protostars (e.g., Yen et al.~2013), and (2) protostars having higher masses tend to exhibit larger Keplerian disks (e.g., Harsono et al.~2014; Aso et al.~2015). 
These results support a scenario where gradually more angular momentum is transferred to the inner envelope and disk-forming region as collapse proceeds. 
However, the efficiency of the angular momentum transfer remains unclear. 
Around a few Class 0 and I protostars, the radial profiles of the rotational velocity in the protostellar envelopes on scales of 100 to 1000 AU are measured to be $\propto r^{-1}$ (Lee et al.~2010; Yen et al.~2013; Ohashi et al.~2014). 
This power-law index is consistent with a fast infalling motion with conserved angular momentum (e.g., Ulrich 1976; Takahashi et al.~2016). 
On the contrary, 
hints of decreasing specific angular momenta on scales from 1000 AU to inner 100 AU have been observed in the Class I protostar HH 111 (Lee et al.~2016) and the Class 0 protostar B335 (Yen et al.~2015b). This can be suggestive of magnetic braking. 
As disk formation and evolution are closely related to the angular momentum transfer in the protostellar envelopes, 
it is essential to probe the gas kinematics from large to small scales and the disk sizes around protostars at different evolutionary stages.

In order to investigate the gas kinematics at an early evolutionary stage and the formation of Keplerian disks,  
we have conducted ALMA observations toward three candidate young protostars, Lupus 3 MMS, IRAS 15398$-$3559, and IRAS 16253$-$2429. They are selected from our SMA sample (Yen et al.~2015a).
These three protostars all have relatively low protostellar masses ($<$0.1 $M_\sun$), inferred from the infalling motions in their protostellar envelopes, 
and they do not show clear signs of a spin-up rotation on a 1000 AU scale, i.e., no signatures of Keplerian disks are seen in our SMA observations (Yen et al.~2015a).
Lupus 3 MMS is a Class 0 protostar with a bolometric luminosity ($L_{\rm bol}$) of 0.41 $L_\sun$ and a bolometric temperature ($T_{\rm bol}$) of 39 K in the Lupus 3 cloud at a distance of 200 pc (Tachihara et al.~2007; Comer\'{o}n 2008; Dunham et al.~2013).
Our SMA results suggest that the protostellar mass in Lupus 3 MMS can be as low as $<$0.1 $M_\sun$ (Yen et al.~2015a). 
IRAS 15398$-$3559 is a Class 0/I protostar with $L_{\rm bol}$ of 1.2 $L_\sun$ and $T_{\rm bol}$ of 61 K in the Lupus 1 cloud at a distance of 150 pc (Froebrich 2005; Comer\'{o}n 2008). 
Early single-dish observations of its CO outflow suggest IRAS 15398$-$3559 is close to face on (van Kempen et al.~2009). 
Recent SMA and ALMA observations show that it is actually closer to edge on (Oya et al.~2014; Bjerkeli et al.~2016). 
With this new estimated inclination angle ($\sim$70\arcdeg), 
our SMA data suggest a low protostellar mass ($<$0.1 $M_\sun$) and a low specific angular momentum in the protostellar envelope ($\sim$1 $\times$ 10$^{-4}$ km s$^{-1}$ pc; Yen et al.~2015a).
IRAS 16253$-$2429 is a Class 0 protostar with $L_{\rm bol}$ of 0.24 $L_\sun$ and $T_{\rm bol}$ of 36 K in the $\rho$ Ophiuchus star-forming region at a distance of 125 pc (Dunham et al.~2013). 
Both CARMA and our SMA results suggest that its protostellar mass is $<$0.1 $M_\sun$ (Tobin et al.~2012b; Yen et al.~2015a). 
These three protostars are all embedded in dense cores with masses $\gtrsim$0.5 $M_\sun$ (Froebrich 2005; Tachihara et al.~2007; Enoch et al.~2009).
Therefore, they are excellent targets to study the gas motions on a 100 AU scale at an early evolutionary stage. 

In the present paper, we report our observational results of Lupus 3 MMS, IRAS 15398$-$3559, and IRAS 16253$-$2429, in the 1.3 mm continuum and the $^{12}$CO (2--1; 230.538 GHz), C$^{18}$O (2--1; 219.560358 GHz), and SO (5$_6$--$4_5$; 219.949433 GHz) lines obtained with ALMA.
The paper is organized as follows: Section 2 describes the details of the observations. Section 3 presents the overall observational results. Section 4 discusses the orientation and inclination of the outflows using the $^{12}$CO emission, the rotational profiles of the protostellar envelopes using the C$^{18}$O and SO emission, and our kinematic models for the C$^{18}$O emission. Section 5 explores disk formation and the emerging evolutionary trend of disk sizes based on our results together with those from the literature. 

\floattable
\begin{deluxetable}{lccccccccccc}
\tablenum{1}
\label{obs}
\tablewidth{0pt}
\tablecaption{Summary of Imaging Parameters}
\tablehead{Source & \multicolumn{2}{c}{1.3 mm continuum} && \multicolumn{2}{c}{$^{12}$CO (2--1)} && \multicolumn{2}{c}{C$^{18}$O (2--1)} && \multicolumn{2}{c}{SO (5$_6$--4$_5$)} \\
 \cline{2-3}  \cline{5-6} \cline{8-9}  \cline{11-12} 
& beam & noise && beam & noise && beam & noise && beam & noise}
\startdata
Lupus 3 MMS & 0\farcs49 $\times$ 0\farcs46 ($8\arcdeg$) & 0.2 && 0\farcs51 $\times$ 0\farcs46 ($24\arcdeg$) & 2.5 && 0\farcs53 $\times$ 0\farcs48 ($21\arcdeg$) & 2.9 && 0\farcs53 $\times$ 0\farcs47 ($20\arcdeg$) & 3.5 \\
IRAS 15398$-$3559 & 0\farcs49 $\times$ 0\farcs48 ($44\arcdeg$) & 0.03 && 0\farcs51 $\times$ 0\farcs47 ($44\arcdeg$) & 2.6 && 0\farcs53 $\times$ 0\farcs49 ($42\arcdeg$) & 3 && 0\farcs53 $\times$ 0\farcs48 ($39\arcdeg$) & 3.7 \\
IRAS 16253$-$2429 & 1\farcs12 $\times$ 0\farcs73 ($80\arcdeg$) & 0.03 && 1\farcs15 $\times$ 0\farcs73 ($82\arcdeg$) & 2.2 && 1\farcs19 $\times$ 0\farcs77 ($81\arcdeg$) & 2.5 && 1\farcs2 $\times$ 0\farcs76 ($81\arcdeg$) & 3 \\
\enddata
\tablecomments{Noise of the 1.3 mm continuum images is in units of mJy Beam$^{-1}$, for molecular-line images it is in units of mJy Beam$^{-1}$ per channel. 
The channel width is 0.5 km s$^{-1}$ for $^{12}$CO (2--1) and 0.17 km s$^{-1}$ for C$^{18}$O (2--1) and SO (5$_6$--4$_5$).}
\end{deluxetable}

\section{Observations}
ALMA cycle-2 observations toward Lupus 3 MMS and IRAS 15398$-$3559 were conducted on April 30, 2014 with 34 antennas and on May 19, 2014 and June 6, 2014 with 36 antennas.  
J1427$-$4206, Titan, and J1534$-$3526 were observed as bandpass, flux, and phase calibrators, respectively. 
The total integration time on Lupus 3 MMS and 15398$-$3559 was 88 min each.
The shortest baseline length was 13 k$\lambda$, and the longest was 498 k$\lambda$ for Lupus 3 MMS and 478 k$\lambda$ for IRAS 15398$-$3559 at 220 GHz. 
The observations toward IRAS 16253$-$2429 were conducted on January 28, 2015 with 38 antennas and on May 4, 2015 with 35 antennas. 
J1517$-$2422 was observed as a bandpass and flux calibrator (1.93 Jy on Jan 28 and 1.31 Jy on May 4), and J1625$-$2527 as a phase calibrator. 
The total integration time on IRAS 16253$-$2429 was 96 min.
The baseline lengths ranged from 9 k$\lambda$ to 272 k$\lambda$ at 220 GHz.
The typical absolute flux uncertainty of ALMA observations$\footnotemark[1]$ at 1 mm wavelength is 10\%.
In all of these observations, 
the correlator was configured in the Frequency Division Mode. 
Two spectral windows, each with a bandwidth of 1875 MHz, were assigned to the 1.3 mm continuum. 
Four spectral windows, each with a bandwidth of 58.6 MHz, were set to the C$^{18}$O (2--1), SO (5$_6$--4$_5$), $^{13}$CO (2--1), and N$_2$D$^+$ (3--2) lines with 960 spectral channels in each window. 
One spectral window with a bandwidth of 117.2 MHz was assigned to the $^{12}$CO (2--1) line with 1920 channels. 
The channel width in all the spectral windows of the molecular lines was 122 kHz. 
Calibration of the raw visibility data was performed with the standard reduction script for the cycle-2 data, which uses tasks in Common Astronomy Software Applications (CASA), and without self-calibration.
The calibrated visibility data of the 1.3 mm continuum and the molecular-line emission were Fourier-transformed with the Briggs robust parameter of +0.5 and CLEANed with the CASA task ``clean''. 
The image cube of the $^{12}$CO (2--1) line was made at a velocity resolution of 0.5 km s$^{-1}$, and those of the C$^{18}$O (2--1) and SO (5$_6$--4$_5$) lines at a resolution of 0.17 km s$^{-1}$. 
The angular resolutions and the noise levels of all the images are listed in Table \ref{obs}.

\footnotetext[1]{Described in the ALMA Proposer's Guide on the website for the ALMA Science Portal.}

\floattable
\begin{deluxetable}{lccccccc}
\tablenum{2}
\label{conti}
\tablewidth{0pt}
\tablecaption{Gaussian Fitting Results of 1.3 mm Continuum Emission}
\tablehead{ & \multicolumn{3}{c}{Extended} & & \multicolumn{3}{c}{Compact} \\
\cline{2-4} \cline{6-8} 
Source & Flux & Deconvolved Size & $M_{\rm 1.3mm}$ & & Flux & Deconvolved Size & $M_{\rm 1.3mm}$ \\
 & (mJy) & & ($M_\sun$) & & (mJy) & & ($M_\sun$)}
\startdata
Lupus 3 MMS & \nodata & \nodata & \nodata & & 185.1 & 0\farcs41 $\times$ 0\farcs26 (148\arcdeg) & 0.1 \\
IRAS 15398$-$3559 &  161.1 & 8\farcs0 $\times$ 4\farcs2 ($50\arcdeg$) & 7.8 $\times$ 10$^{-2}$ & & 8.3 & point source & 7.2 $\times$ 10$^{-4}$ \\
IRAS 16253$-$2429 & 12.1 & 6\farcs4 $\times$ 3\farcs5 ($121\arcdeg$) & 4.3 $\times$ 10$^{-3}$ & & 13.8 & 0\farcs26 $\times$ 0\farcs18 (112\arcdeg) & 2.9 $\times$ 10$^{-3}$ \\
\enddata
\tablecomments{The uncertainties of the fitted fluxes and sizes are less than 1\%, and those in the position angles are less than 1$\arcdeg$ in Lupus 3 MMS and IRAS 15398$-$3559 and 2$\arcdeg$ in IRAS 16253$-$2429. The absolute flux uncertainty is 10\%.}
\end{deluxetable}

\section{Results}

\begin{figure*}
\figurenum{1}
\plotone{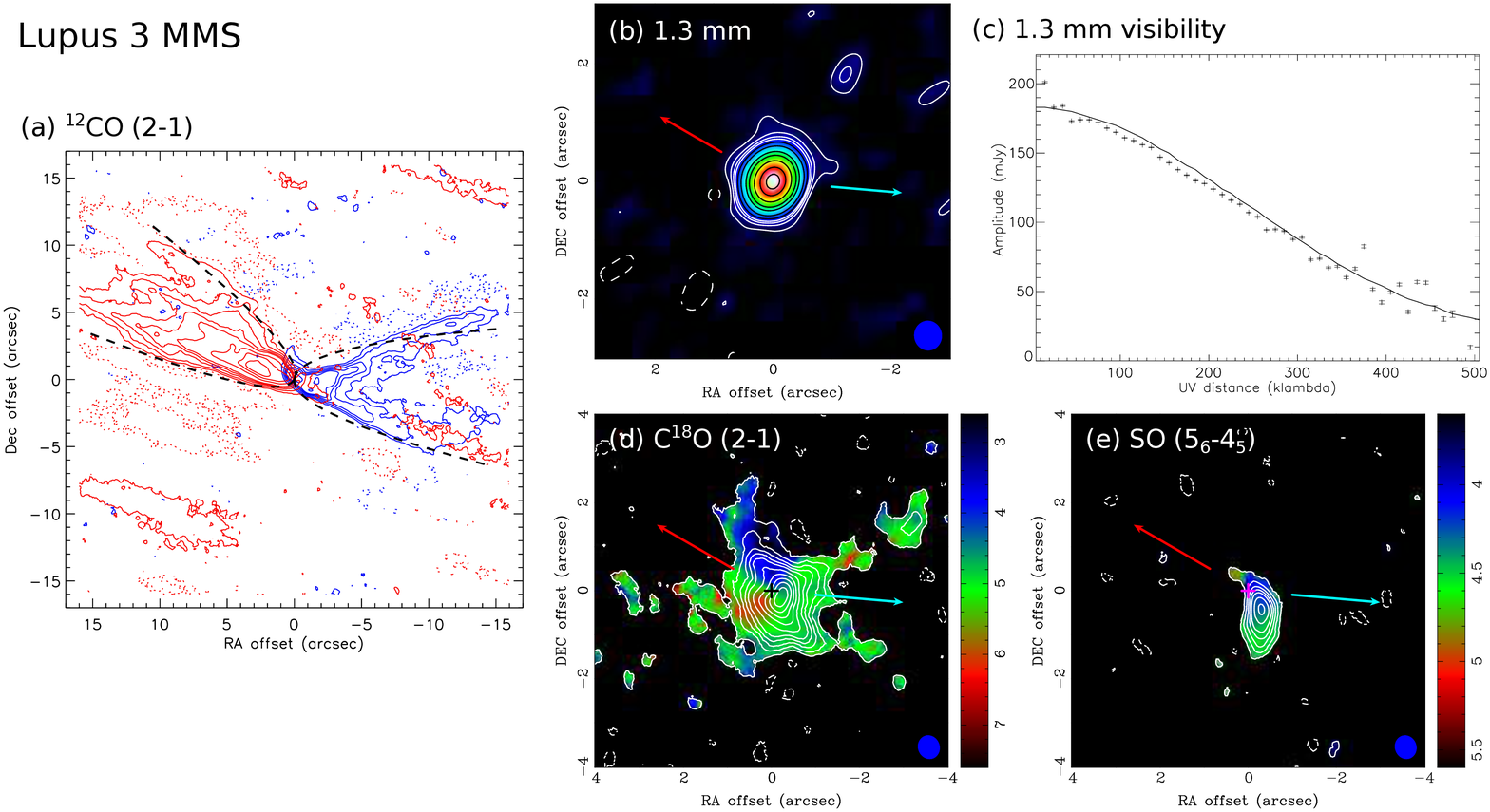}
\caption{ALMA observational results of Lupus 3 MMS. (a) Moment 0 map of blue- and red-shifted $^{12}$CO (2--1). Black dashed lines delineate the best-fit projected parabolic shapes to the observed $^{12}$CO outflow morphologies. Contour levels start from 5$\sigma$ in steps of powers of two, i.e., 5$\sigma$, 10$\sigma$, 20$\sigma$......, where 1$\sigma$ is 4.5 and 5.9 mJy beam$^{-1}$ km s$^{-1}$ in the blue- and red-shifted range. (b) 1.3 mm continuum. Contour levels are 5$\sigma$, 10$\sigma$, 15$\sigma$, 20$\sigma$, 30$\sigma$, 50$\sigma$, 100$\sigma$, 150$\sigma$, 250$\sigma$, 400$\sigma$, 550$\sigma$, where 1$\sigma$ is 0.2 mJy Beam$^{-1}$. (c) Visibility amplitude as a function of $uv$ distance of the 1.3 mm continuum emission. The solid line presents the visibility amplitude profile of the fitted 2-dimensional Gaussian. (d) \& (e) Moment 0 maps (contour) overlaid on moment 1 maps (color) of C$^{18}$O and SO. Moment 1 maps (color bars) are in units of km s$^{-1}$. Contour levels are from 3$\sigma$ in steps of 3$\sigma$ to 15$\sigma$ and then in steps of 5$\sigma$, where 1$\sigma$ is 2.8 and 1.8 mJy beam$^{-1}$ km s$^{-1}$ in C$^{18}$O and SO, respectively. Blue and red arrows represent the axes of the blue- and red-shifted $^{12}$CO outflows. Crosses denote the protostellar position. Filled blue ellipses show the sizes of the synthesized beams.}\label{lupus3mms}
\end{figure*}

\subsection{Lupus 3 MMS}
Figure \ref{lupus3mms} presents our observational results of Lupus 3 MMS with ALMA.
The blue- and red-shifted $^{12}$CO (2--1) emission shows V-shaped (or fan-like) structures with the apices located at the center (Fig.~\ref{lupus3mms}a), 
most likely tracing the wall of the outflow cavity.
The blueshifted outflow is oriented toward the west, and the redshifted outflow toward the northeast. 
This orientation is consistent with that of the $^{12}$CO outflow observed with ASTE (Dunham et al.~2014) and the SMA (Yen et al.~2015a) at lower angular resolutions.
It is also consistent with the infrared image of the outflow cavity and the direction of the Herbig-Haro object HH 78 that is located in Lupus 3 (Nakajima et al.~2003; Tachihara et al.~2007). 
The 1.3 mm continuum emission shows a compact component with an apparent size of $\sim$2$\arcsec$ elongated along the northwest--southeast direction (Fig.~\ref{lupus3mms}b). 
Its visibility amplitude profile as a function of $uv$ distance can be represented by a single Gaussian component.
By fitting the visibility data with a 2-dimensional Gaussian function with the CASA task {\it uvmodelfit}, 
the peak position is measured to be $\alpha$(J2000) = $16^{\rm h}09^{\rm m}18\fs09$, $\delta$(J2000) = $-39\arcdeg04\arcmin53\farcs3$, 
and the total integrated flux, full-width-half-maximum (FWHM) size, and position angle of the Gaussian component are estimated to be 185.1 mJy, 0\farcs41 $\times$ 0\farcs26 (82 AU $\times$ 52 AU), and 148\arcdeg, respectively (Table \ref{conti}). 
In the following, this peak position is adopted as the protostellar position of Lupus 3 MMS.
The orientation of the continuum emission is perpendicular to the redshifted outflow, 
and it likely traces a disk-like or flattened structure around the protostar. 
The mass of the circumstellar material traced by the 1.3 mm continuum emission ($M_{\rm 1.3mm}$) can be estimated as
\begin{equation}\label{md}
M_{\rm 1.3mm} = \frac{F_{\rm 1.3mm}d^{2}} {\kappa_{\rm 1.3 mm} B(T_{\rm dust})},
\end{equation}
where $F_{\rm 1.3mm}$ is the total integrated 1.3 mm flux, $d$ is the distance to the source, $\kappa_{\rm 1.3 mm}$ is the dust mass opacity at 1.3 mm, $T_{\rm dust}$ is the dust temperature, and $B(T_{\rm dust})$ is the Planck function at a temperature of $T_{\rm dust}$. 
The typical number density in protostellar envelopes on a scale of hundreds of AU is on the order of 10$^6$ cm$^{-3}$. 
Hence, we adopt $\kappa_{\rm 1.3 mm} = 0.009$ cm$^2$ g$^{-1}$ from the dust coagulation model of the MRN (Mathis, Rumpl, \& Nordsieck 1977) distribution with thin ice mantles at a number density of 10$^6$ cm$^{-3}$ computed by Ossenkopf and Henning (1994), and we assume a gas-to-dust mass ratio of 100.
The deconvolved FWHM size of the 1.3 mm continuum component is $\sim$80 AU.
With a typical $T_{\rm dust}$ of 30 K on a 100 AU scale, $M_{\rm 1.3mm}$ is estimated to be 0.1 $M_\sun$. 
On the contrary, 
if we adopt the same frequency function for dust mass opacity with $\beta = 1$ as Beckwith et al.~(1990), 
$\kappa_{\rm 1.3 mm}$ is a factor of two larger, 0.02 cm$^2$ g$^{-1}$, 
that results in a factor of two smaller estimated mass. 
In addition, the estimated $M_{\rm 1.3mm}$ is inversely proportional to $T_{\rm dust}$ with the Rayleigh--Jeans limit.

The C$^{18}$O emission shows a compact component with an apparent size of $\sim$2$\arcsec$ around the protostar and extensions with lengths of 2\arcsec--4$\arcsec$ along the cavity wall of the associated outflow (Fig.~\ref{lupus3mms}d). 
Its peak position is located $\sim$0\farcs3 west from the protostellar position. 
This peak offset and the extensions along the outflow cavity wall are likely due to outflow contamination. 
The C$^{18}$O emission exhibits a clear velocity gradient from the northwest (blueshifted) to the southeast (redshifted). 
The direction of the velocity gradient is perpendicular to the redshifted outflow and is along the elongation of the 1.3 mm continuum emission. 
Moreover, the locations of the C$^{18}$O blue- and red-shifted emission are symmetric with respect to the protostellar position. 
Hence, the C$^{18}$O emission likely traces the dominant rotation of the inner envelope around the protostar. 
In contrast to this, 
the SO emission is not centered on the protostellar position, 
and shows an elongated structure toward the south (Fig.~\ref{lupus3mms}e). 
Thus, the SO emission unlikely traces the inner envelope but is more likely related to outflow activities$\footnotemark[2]$.

\footnotetext[2]{SO emission is expected to be enhanced when the dust temperature is above the SO desorption temperature which is $\sim$60 K (e.g., Aikawa et al.~2012).  Thus, SO emission most likely originates from warm regions. Possible heating mechanisms in protostellar sources include protostellar heating, accretion shocks, and outflow shocks. Because of the asymmetric distribution and the offset from the protostar, the origin of the compact SO emission in Lupus 3 MMS is less likely related to protostellar heating or accretion shocks, but is more likely due to outflow activities.}

\begin{figure*}
\figurenum{2}
\plotone{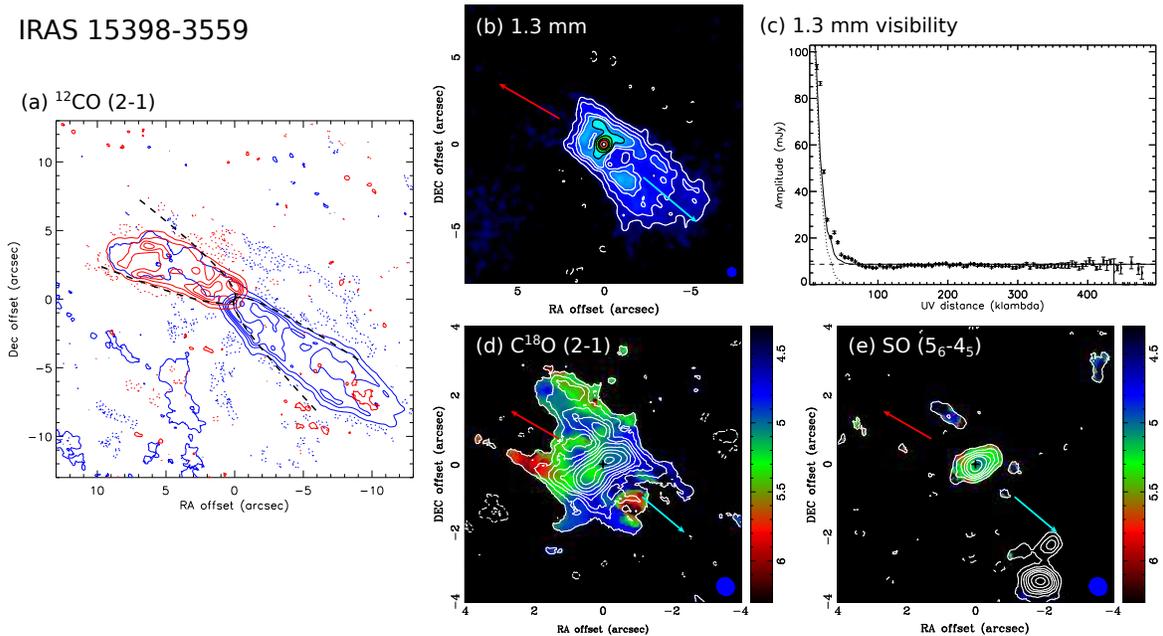}
\caption{Same as Figure \ref{lupus3mms} but for IRAS 15398$-$3559. Dashed and dotted lines in (c) present the visibility amplitude functions of the fitted point source and the extended 2-dimensional Gaussian, respectively, and the solid line shows the summation of the two components. Contour levels in (b) are 5$\sigma$, 10$\sigma$, 15$\sigma$, 20$\sigma$, 30$\sigma$, 50$\sigma$, and then in steps of 50$\sigma$. Others are the same as in Figure \ref{lupus3mms}. 1$\sigma$ is 8 and 6.6 mJy beam$^{-1}$ km s$^{-1}$ in the blue- and red-shifted $^{12}$CO map, it is 0.03 mJy beam$^{-1}$ in the 1.3 mm continuum map, 2 mJy beam$^{-1}$ km s$^{-1}$ in the C$^{18}$O map, and 3.4 mJy beam$^{-1}$ km s$^{-1}$ in the SO map.}\label{i15398}
\end{figure*}

\subsection{IRAS 15398$-$3559}
Figure \ref{i15398} presents our observational results of IRAS 15398$-$3559 with ALMA.
The blue- and red-shifted $^{12}$CO (2--1) emission is oriented toward southwest and northeast, respectively (Fig.~\ref{i15398}a). 
This orientation is consistent with that of the outflow observed in $^{12}$CO (2--1) with the SMA (Yen et al.~2015a; Bjerkeli et al.~2016) and in the H$_2$CO and CCH lines with ALMA (Oya et al.~2014). 
The outflow morphology is more collimated than that of Lupus 3 MMS and IRAS 16253$-$2429, 
and it does not exhibit simple V-shaped or fan-like structures. 
Furthermore, the heads of the outflow 
display bow-like features. 
The 1.3 mm continuum is composed of a brighter compact component with an apparent size of $\sim$1$\arcsec$ at the center and an extended component along the blueshifted outflow (Fig.~\ref{i15398}b).
The visibility amplitude profile of the 1.3 mm continuum emission shows a quick drop from a $uv$ distance around 10 k$\lambda$ to 50 k$\lambda$, 
and becomes then almost constant up to 480 k$\lambda$ (Fig.~\ref{i15398}c). 
This profile can be represented with a 2-dimensional Gaussian function and a point source. 
By fitting the visibility data with these two components, 
the position of the point source is measured to be $\alpha$(J2000) = $15^{\rm h}43^{\rm m}02\fs24$, $\delta$(J2000) = $-34\arcdeg09\arcmin06\farcs8$, 
with a flux of 8.3 mJy (Table \ref{conti}). 
This position is adopted as the protostellar position of IRAS 15398$-$3559.
The extended component has a deconvolved FWHM size of 8$\arcsec$ $\times$ 4$\farcs$2 (1200 AU $\times$ 600 AU) and a total integrated flux of 161.1 mJy (Table \ref{conti}). 
Since the central component is unresolved, we adopt the beam size as the upper limit of its size, which corresponds to a radius of 35 AU. 
Although typically the temperature at a radius of 35 AU in protostellar envelopes is 20--50 K (e.g., Shirley et al.~2000), 
our kinematic model suggests a relatively high temperature $\gtrsim$100 K on a 100 AU scale around IRAS 15398$-$3559 (Section \ref{model}). 
In addition, J{\o}rgensen et al.~(2013) suggest, based on their ALMA results in the C$^{17}$O, H$^{13}$O$^+$ and CH$_3$OH lines, that there was an accretion burst during the last 100--1000 yr in IRAS 15398$-$3559, resulting in an extended warm (100 K) region of a 100 AU scale. 
Thus, $M_{\rm 1.3mm}$ of the central point source is estimated with $T_{\rm dust}$ of 100 K to be 7.2 $\times$ 10$^{-4}$ $M_\sun$ (Equation \ref{md}).
For the extended component, a lower $T_{\rm dust}$ of 20 K, a typical temperature on a scale of hundreds of AU in protostellar envelopes, is adopted.
Its $M_{\rm 1.3mm}$ is then estimated to be 7.8 $\times$ 10$^{-2}$ $M_\sun$ (Table \ref{conti}).

The C$^{18}$O emission shows a compact component with an apparent size of $\sim$2$\arcsec$ elongated along the northwest--southeast direction centered on the protostellar position (Fig.~\ref{i15398}d). 
Besides, the extended structures delineating the outflow cavity wall are clearly seen. 
The compact component exhibits a velocity gradient along its elongation, where the blueshifted emission extends toward northwest and the redshifted emission to the southeast. 
On the other hand, the SO emission shows a compact component with an apparent size of $\sim$1$\arcsec$ at the center, and its peak position coincides with the protostellar position (Fig.~\ref{i15398}e). 
An additional SO component is located to the southwest along the direction of the blueshifted outflow. 
The central SO component likely traces the inner warm envelope, while the southwestern component is likely related to  outflow activities. 
There is no clear velocity feature seen in the intensity-weighted mean velocity (moment 1) map of the SO compact component around the protostar.

\begin{figure*}
\figurenum{3}
\plotone{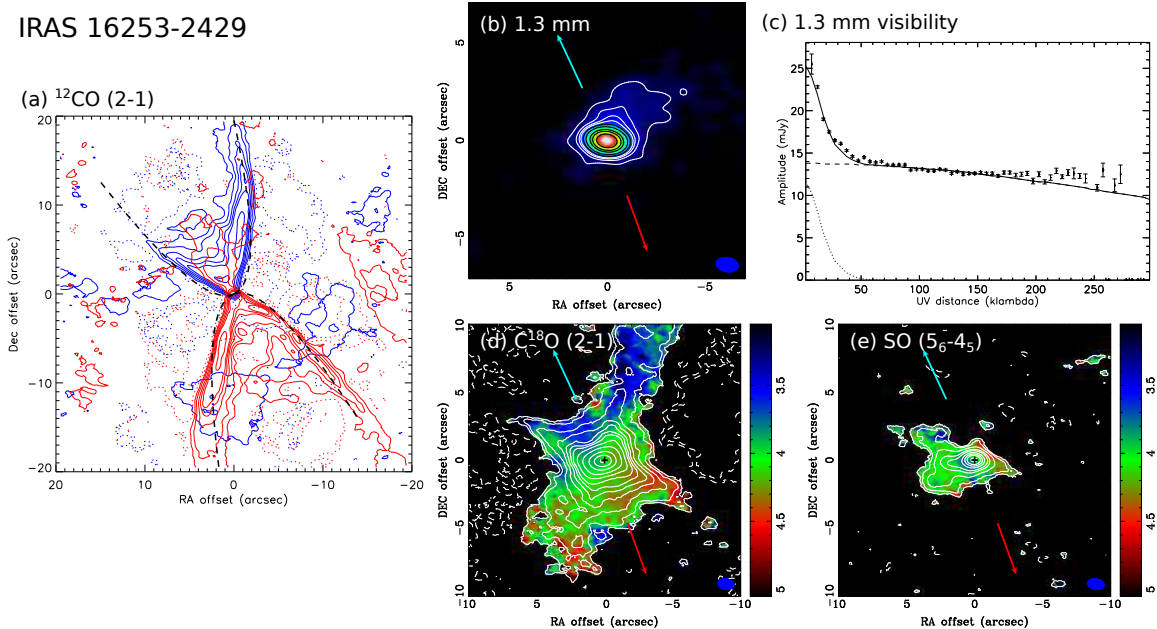}
\caption{Same as Figure \ref{i15398} but for IRAS 16253$-$2429. Dashed and dotted lines in (c) present the visibility amplitude functions of the fitted compact and the extended 2-dimensional Gaussian, respectively, and the solid line shows the summation of the two components. Contour levels in (d) are 3$\sigma$, 5$\sigma$, 10$\sigma$, 15$\sigma$, 20$\sigma$, 25$\sigma$, 30$\sigma$, 40$\sigma$, 60$\sigma$, 100$\sigma$, and then in steps of 50$\sigma$. Those in (e) are 3$\sigma$, 5$\sigma$, 10$\sigma$, 20$\sigma$, 30$\sigma$ ,50$\sigma$, 70$\sigma$, 90$\sigma$, 120$\sigma$, 150$\sigma$ and then in steps of 50$\sigma$. Others are the same as in Figure \ref{i15398}. 1$\sigma$ is 6 and 5.3 mJy beam$^{-1}$ km s$^{-1}$ in the blue- and red-shifted $^{12}$CO map, it is 0.03 mJy beam$^{-1}$ in the 1.3 mm continuum map, 2.4 mJy beam$^{-1}$ km s$^{-1}$ in the C$^{18}$O moment 0 map, and 2.2 mJy beam$^{-1}$ km s$^{-1}$ in the SO moment 0 map.}\label{i16253}
\end{figure*}

\subsection{IRAS 16253$-$2429}
Figure \ref{i16253} presents our observational results of IRAS 16253$-$2429 with ALMA. 
The blue- and red-shifted $^{12}$CO (2--1) emission shows V-shaped structures with the apices at the center, oriented along a northeast--southwest direction (Fig.~\ref{i16253}a). 
This orientation is consistent with that of the $^{12}$CO (3--2) outflow observed with the JCMT (van der Marel et al.~2013) and the reflection nebula observed in infrared (Tobin et al.~2010). 
The 1.3 mm continuum shows a central compact component with an apparent size of $\sim$3$\arcsec$ and an elongated structure with an apparent size of $\sim$5$\arcsec$ extending toward northwest (Fig.~\ref{i16253}b). 
The visibility amplitude profile shows a quick drop from a $uv$ distance around 10 k$\lambda$ to 50 k$\lambda$, 
and then it flattens (Fig.~\ref{i16253}c). 
This profile can be represented by fitting two 2-dimensional Gaussians. 
In this way, the peak position of the compact component is measured to be $\alpha$(J2000) = $16^{\rm h}28^{\rm m}21\fs62$, $\delta$(J2000) = $-24\arcdeg36\arcmin24\farcs2$. 
In the following, this position is adopted as the protostellar position of IRAS 16253$-$2429. 
The total integrated flux, FWHM size, and position angle of the compact and extended Gaussian component are estimated to be 13.8 mJy, 0\farcs26 $\times$ 0\farcs18 (33 AU $\times$ 22 AU), and 112$\arcdeg$, and 12.1 mJy, 6\farcs4 $\times$ 3\farcs5 (800 AU $\times$ 437 AU), and 121$\arcdeg$, respectively (Table \ref{conti}). 
Both the compact and the extended continuum component are oriented perpendicularly to the outflow axis. 
The observed visibility amplitude profile shows some flux excess at $uv$ distances beyond 170 k$\lambda$, 
as compared to the fitted Gaussians. This suggests that the innermost structure can be still more compact than the derived Gaussian component. 
With Equation \ref{md} and a typical $T_{\rm dust}$ of 30 K on a 100 AU scale in protostellar envelopes, $M_{\rm 1.3mm}$ of the compact component is estimated to be 2.9 $\times$ 10$^{-3}$ $M_\sun$, 
while the extended component is 4.3 $\times$ 10$^{-3}$ $M_\sun$ assuming its $T_{\rm dust}$ is 20 K.

The C$^{18}$O emission reveals a compact component with an apparent size of $\sim$5$\arcsec$, centered on the protostellar position (Fig.~\ref{i16253}d). 
In addition, there are extensions along the outflow cavity wall in the C$^{18}$O emission. 
A clear velocity gradient from northeast (blueshifted) to southwest (redshifted), identical to the orientation of the $^{12}$CO outflow, is seen in the C$^{18}$O emission. 
Hence, the outer extensions of the C$^{18}$O emission are likely related to the outflow. 
This northeast--southwest velocity gradient is not seen in the central compact component, 
suggesting the gas motion in the inner region is not dominated by the outflow.
The SO emission shows a compact component with an apparent size of $\sim$1$\arcsec$ with its peak position at the protostellar position. 
Extended SO structures are oriented toward the east (Fig.~\ref{i16253}e).
These extended structures are likely along the outflow cavity and related to outflow activities. 
The central component displays a clear velocity gradient from the southeast (blueshifted) to the northwest (redshifted), perpendicular to the outflow axis. 
The southeast--northwest velocity gradient in the SO emission 
is likely tracing the rotation of the inner warm envelope. 

\section{Analysis}

\floattable
\begin{deluxetable}{lccccccccc}
\tablenum{3}
\label{outflowfit}
\tablewidth{0pt}
\tablecaption{Estimated Outflow Orientations and Inclinations}
\tablehead{ & \multicolumn{4}{c}{Redshifted Lobe} & & \multicolumn{4}{c}{Blueshifted Lobe} \\
\cline{2-5} \cline{7-10}
Source & PA & $i$ & $c_0$ & $v_0$ & & PA & $i$ & $c_0$ & $v_0$ \\
& & & (arcsec$^{-1}$) & (km s$^{-1}$ arcsec$^{-1}$) & & & &  (arcsec$^{-1}$) & (km s$^{-1}$ arcsec$^{-1}$)}
\startdata
Lupus 3 MMS & 60$\pm$5\arcdeg & 65$\pm$5\arcdeg & 0.8$\pm$0.1 & 1.7$^{+0.2}_{-0.3}$ & & 265$\pm$5\arcdeg & 50$\pm$20\arcdeg & 0.7$^{+0.4}_{-0.1}$ & 1.0$\pm$0.5\\
IRAS 15398$-$3559 & 60$\pm$5\arcdeg & 70$\pm$5\arcdeg & 1.4$\pm$0.1 & 4.8$\pm$0.4 & &  230$\pm$5\arcdeg & 70$\pm$5\arcdeg & 1.7$\pm$0.1 & 5.0$^{+0}_{-1.3}$ \\
IRAS 16253$-$2429 & 200$\pm$5\arcdeg & 60$\pm$5\arcdeg & 0.3 & 1.8$^{+0.6}_{-0}$ & & 25$\pm$5\arcdeg & 65$\pm$5\arcdeg & 0.3 & 3.1$^{+0.5}_{-0.6}$ \\
\enddata
\tablecomments{The errors of $i$, $c_0$, and $v_0$ here only represent the degeneracy between the parameters but not measurement errors. The difference between the projected velocity structures in our outflow models with the parameters within the errors is less than 15\%. On the contrary, PA can be robustly measured with the uncertainties of 5\arcdeg.}
\end{deluxetable}

\begin{figure*}
\figurenum{4}
\plotone{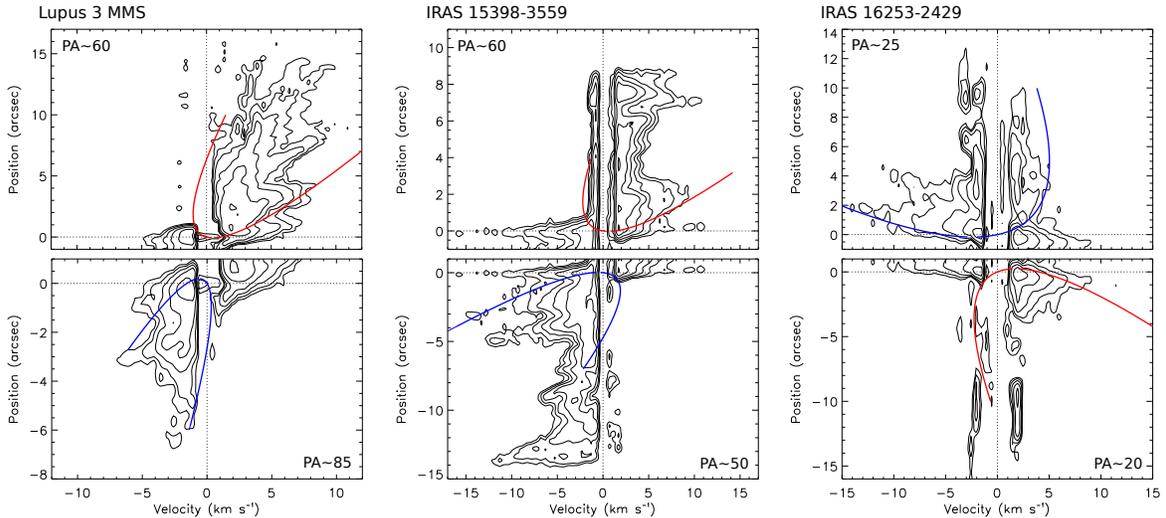}
\caption{P--V diagrams of the $^{12}$CO emission along the blue- and red-shifted outflow axes in Lupus 3 MMS, IRAS 15398$-$3559, and IRAS 16253$-$2429 observed with ALMA. The position angles $PA$ of the outflow axes are shown in the upper-left and bottom-right corner in each diagram. Contour levels start from 5$\sigma$ in steps of powers of two, i.e., 5$\sigma$, 10$\sigma$, 20$\sigma$......, where the 1$\sigma$ levels are listed in Table \ref{obs}.}\label{copv}
\end{figure*}

\subsection{Outflow Orientations and Inclinations}\label{outflow}
The orientations and inclinations of the outflows in these three protostars have been studied with single-dish or SMA observations at lower angular resolutions (van der Marel et al.~2013; Dunham et al.~2014; Yen et al.~2015a; Bjerkeli et al.~2016). 
Our ALMA $^{12}$CO observations have higher angular resolutions and sensitivities and thus, reveal the outflow morphologies and velocity structures more clearly. 
The outflow in IRAS 15398$-$3559 has already been observed with ALMA at comparable angular resolutions of $\sim$0\farcs5 in the emission lines of molecules that have lower abundances as compared to CO, such as H$_2$CO and CCH (Oya et al.~2014).
Our ALMA $^{12}$CO data are adding information of the low-density part of the outflows.
Therefore, we re-estimate the orientations and inclinations of these outflows from our new $^{12}$CO data, adopting the wind-driven-shell model (e.g., Shu et al.~1991, 2000). 
The estimated values are listed in Table \ref{outflowfit}.
These parameters are adopted for our analyses of the gas kinematics in the protostellar envelopes in this present paper. 
 
Our method to compare the observed morphologies and velocity structures of the outflows with the wind-driven outflow model is described in Appendix \ref{wind}.
The best-fit outflow morphologies and velocity structures are shown as black dashed curves in the $^{12}$CO total-integrated intensity (moment 0) maps (Fig.~\ref{lupus3mms}a, \ref{i15398}a, and \ref{i16253}a) and as blue and red solid curves in the $^{12}$CO Position--Velocity (P--V) diagrams (Fig.~\ref{copv}), respectively.
Compared to the previous estimates of outflow inclinations at lower angular resolutions (van der Marel et al.~2013; Dunham et al.~2014; Yen et al.~2015a), 
our results suggest that Lupus 3 MMS and IRAS 16253$-$2429 are less edge-on. 
Bjerkeli et al.~(2016) and Oya et al.~(2014) have also applied the wind-driven outflow model to the outflow in IRAS 15398$-$3559, observed in $^{12}$CO (2--1) with the SMA and in H$_2$CO and CCH with ALMA. 
Our estimated inclination angle is consistent with theirs, although $c_0$ and $v_0$, the two fitting parameters that describe morphology and velocity structures, are different by a factor of a few. 
This difference can be due to (1) $^{12}$CO tracing lower-density parts of the outflow as compared to H$_2$CO and CCH, and (2) the different methods comparing the observations with Equation \ref{windeq}. 
In our method, we search for the best $c_0$ and $v_0$ to generate curves delineating the 5$\sigma$ contours, 
while they generate curves passing through the contour ridges. 
However, the key parameter for the analyses of the gas kinematics in the protostellar envelopes in the present paper is the inclination angle, and the estimates of the inclination angles appear robust 
as these different studies derive consistent values.
In addition, 
in Lupus 3 MMS and IRAS 16253$-$2429, the inner compact components of the 1.3 mm continuum are resolved. 
Assuming these components trace the geometrically-thin circular disks, 
their inclination angles can also be estimated from the aspect ratios of the major and minor axes as $\arccos$(minor/major).
Our estimates of the inclination angles of 60$\arcdeg$ in Lupus 3 MMS and IRAS 16253$-$2429 are comparable to those from the aspect ratios of the 1.3 mm continuum emission ($\sim$50$\arcdeg$).
We further note that the V-shaped morphology of the blue- and red-shifted $^{12}$CO emission in IRAS 16253$-$2429 is interpreted as two precessing bipolar jets from a binary system in Hsieh et al.~(2016). 
Their model of precessing jets also suggests an inclination angle of 60$\arcdeg$--80$\arcdeg$, comparable to our estimate. 

\begin{figure*}
\figurenum{5}
\plotone{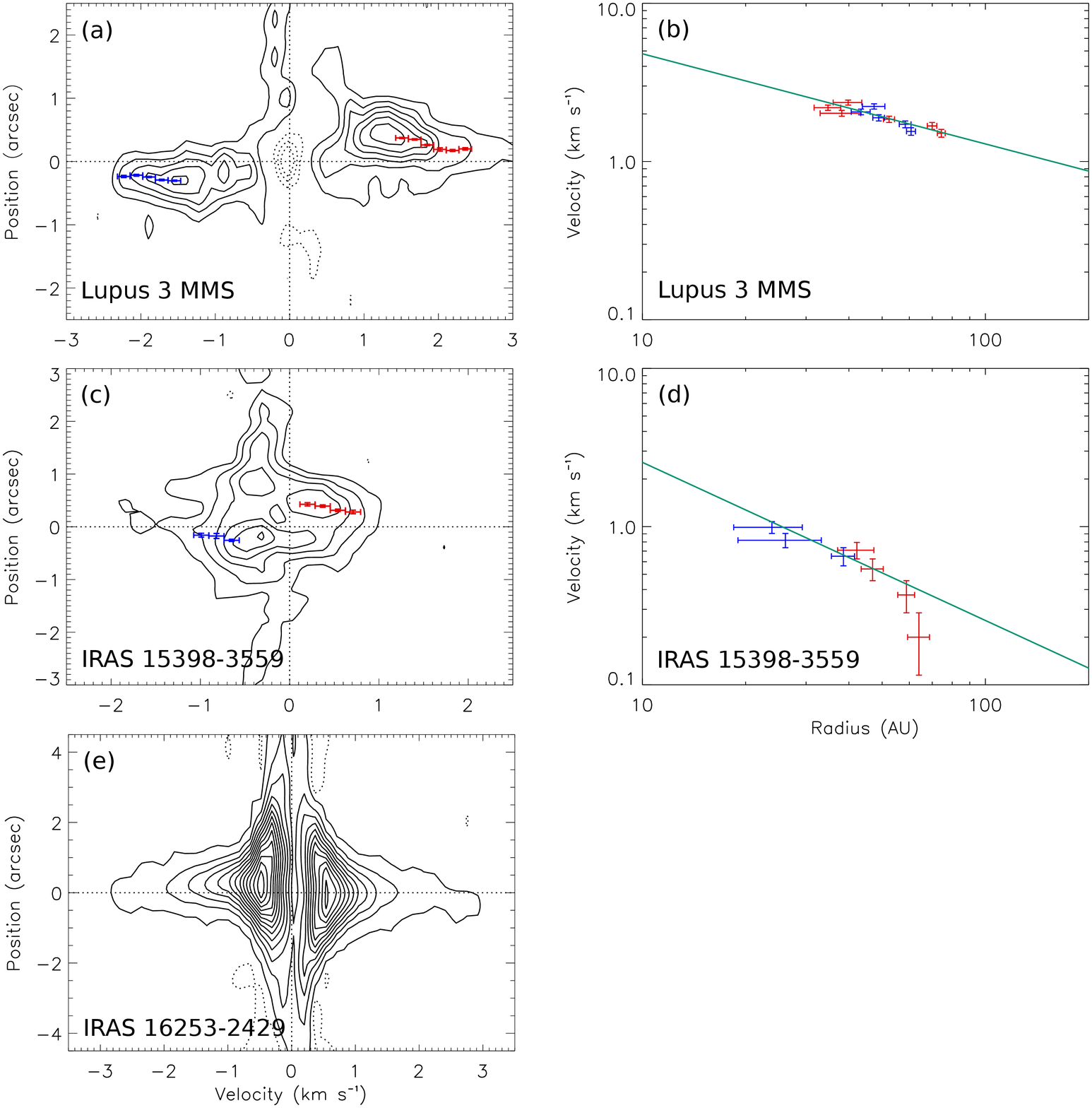}
\caption{P--V diagrams of the C$^{18}$O emission along the major axis in (a) Lupus 3 MMS, (c) IRAS 15398$-$3559, and (e) IRAS 16253$-$2429 observed with ALMA. Blue and red data points denote the measured $R_{\rm rot}$ and $V_{\rm rot}$. Contour levels all start from 3$\sigma$, and are in steps of 3$\sigma$ in (a) and (c) and 10$\sigma$ in (e), where the 1$\sigma$ levels are listed in Table \ref{obs}. (b) and (d) present the rotational profiles from the data points in the P--V diagrams. Solid lines denote the fitted power-law functions.}\label{c18opv}
\end{figure*}

\subsection{Rotational Profiles}\label{Rot}
We follow the same method as in Yen et al.~(2013) to measure rotational profiles, $V_{\rm rot} \propto {R_{\rm rot}}^f$, where $V_{\rm rot}$ and $R_{\rm rot}$ are rotational velocities and radii, and $f$ is the power-law index.
Additionally, we include the systemic velocity ($V_{\rm sys}$) as a free parameter. 
The method is briefly summarized in Appendix \ref{vr} and described in detail in Yen et al.~(2013). 

\begin{figure*}
\figurenum{6}
\plotone{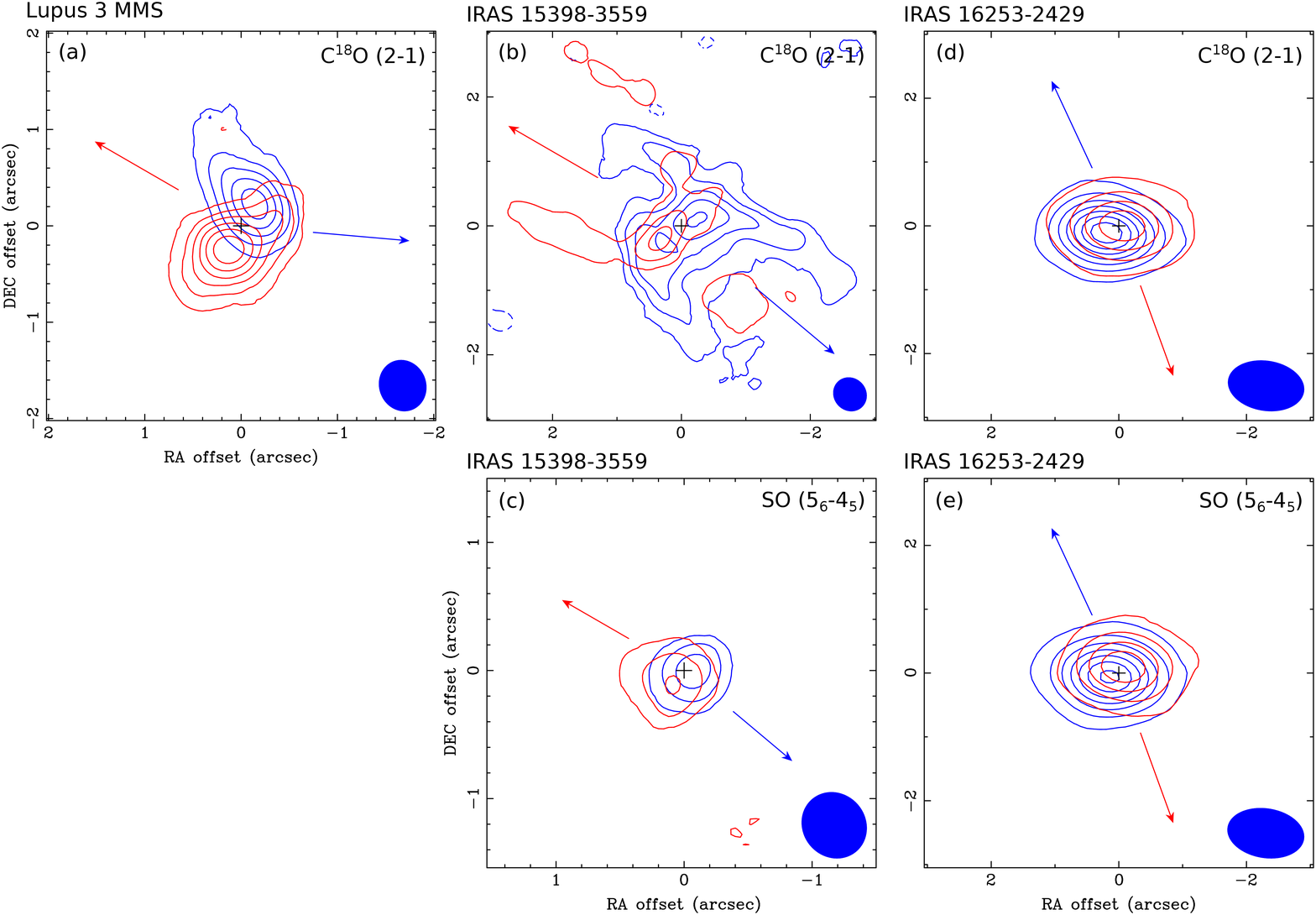}
\caption{Moment 0 maps of the high-velocity blue- and red-shifted C$^{18}$O and SO emission in Lupus 3 MMS, IRAS 15398$-$3559, and IRAS 16253$-$2429. Blue and red arrows indicate the directions of the blue- and red-shifted outflows, and crosses denote the protostellar positions. Filled blue ellipses show the sizes of the synthesized beams. The integrated velocity ranges are $\Delta V \gtrsim$ 1.4 km s$^{-1}$ in (a), $\gtrsim$ 0.3 km s$^{-1}$ in (b), $\gtrsim$ 1 km s$^{-1}$ in (c), $\gtrsim$ 0.4 km s$^{-1}$ in (d), and $\gtrsim$ 0.5 km s$^{-1}$ in (e). Contour levels in (a) are from 5$\sigma$ in steps of 5$\sigma$, where 1$\sigma$ is 1.3 and 1.5 mJy Beam$^{-1}$ km s$^{-1}$ in the blue- and red-shifted range, respectively. Those in (b) are from 5$\sigma$ in steps of 5$\sigma$, where 1$\sigma$ is 1.1 and 1 mJy Beam$^{-1}$ km s$^{-1}$ in the blue- and red-shifted range. Those in (c) are from 5$\sigma$ in steps of 3$\sigma$, where 1$\sigma$ is 1.5 and 1.6 mJy Beam$^{-1}$ km s$^{-1}$ in the blue- and red-shifted range. Those in (d) \& (e) are from 5$\sigma$ in steps of 10$\sigma$, where 1$\sigma$ is 1.3 and 1.5 mJy Beam$^{-1}$ km s$^{-1}$ in the blue- and red-shifted C$^{18}$O, and 2 mJy Beam$^{-1}$ km s$^{-1}$ in the blue- and red-shifted SO.}\label{hvmom0}
\end{figure*}

\subsubsection{Lupus 3 MMS}
The rotational profile in Lupus 3 MMS is measured to be $V_{\rm rot} = (0.87\pm0.04) \times (R/R_0)^{-0.57\pm0.03}$ with $V_{\rm sys} = 4.62\pm0.04$ km s$^{-1}$ (Fig.~\ref{c18opv}b), where $R_0$ is 1$\arcsec$ (200 AU in Lupus 3 MMS).
The measured $V_{\rm sys}$ is consistent with the one measured by single-dish observations in the H$^{13}$CO$^+$ (1--0) line (4.6 km s$^{-1}$; Tachihara et al.~2007). 
The power-law index 
is consistent with Keplerian rotation ($f = -0.5$) within uncertainty. This suggests the presence of a Keplerian disk, 
although a possible contamination from the outflow is still seen in the moment 0 maps of the high-velocity C$^{18}$O emission ($\Delta V \gtrsim 1.4$ km s$^{-1}$; Fig.~\ref{hvmom0}a).
No transition from the protostellar envelope to the disk is observed in the rotational profile, which is different from L1527 IRS (Ohashi et al.~2014) and TMC-1A (Aso et al.~2015). 
This could be because the density of the protostellar envelope on a 100 AU scale in Lupus 3 MMS is relatively low as compared to the disk (seen in the visibility amplitude profile of the 1.3 mm continuum as a single Gaussian-like component). 
Thus, the rotational profile of the envelope component is not detected in our observations.
Since the observed rotational velocity follows a $R^{-0.5}$ Keplerian profile up to a radius of $\sim$100 AU, 
the radius of the Keplerian disk is likely at least 100 AU.
Assuming the inclination angle is 60$\arcdeg$ (outflow analysis, Section \ref{outflow}), the observed Keplerian rotation corresponds to a protostellar mass of 0.23 $M_\sun$.
For comparison, the 1.3 mm continuum results suggest a disk mass of $\sim$0.1 $M_\sun$, which is $\sim$40\% of the protostellar mass. 

\begin{figure*}
\figurenum{7}
\plotone{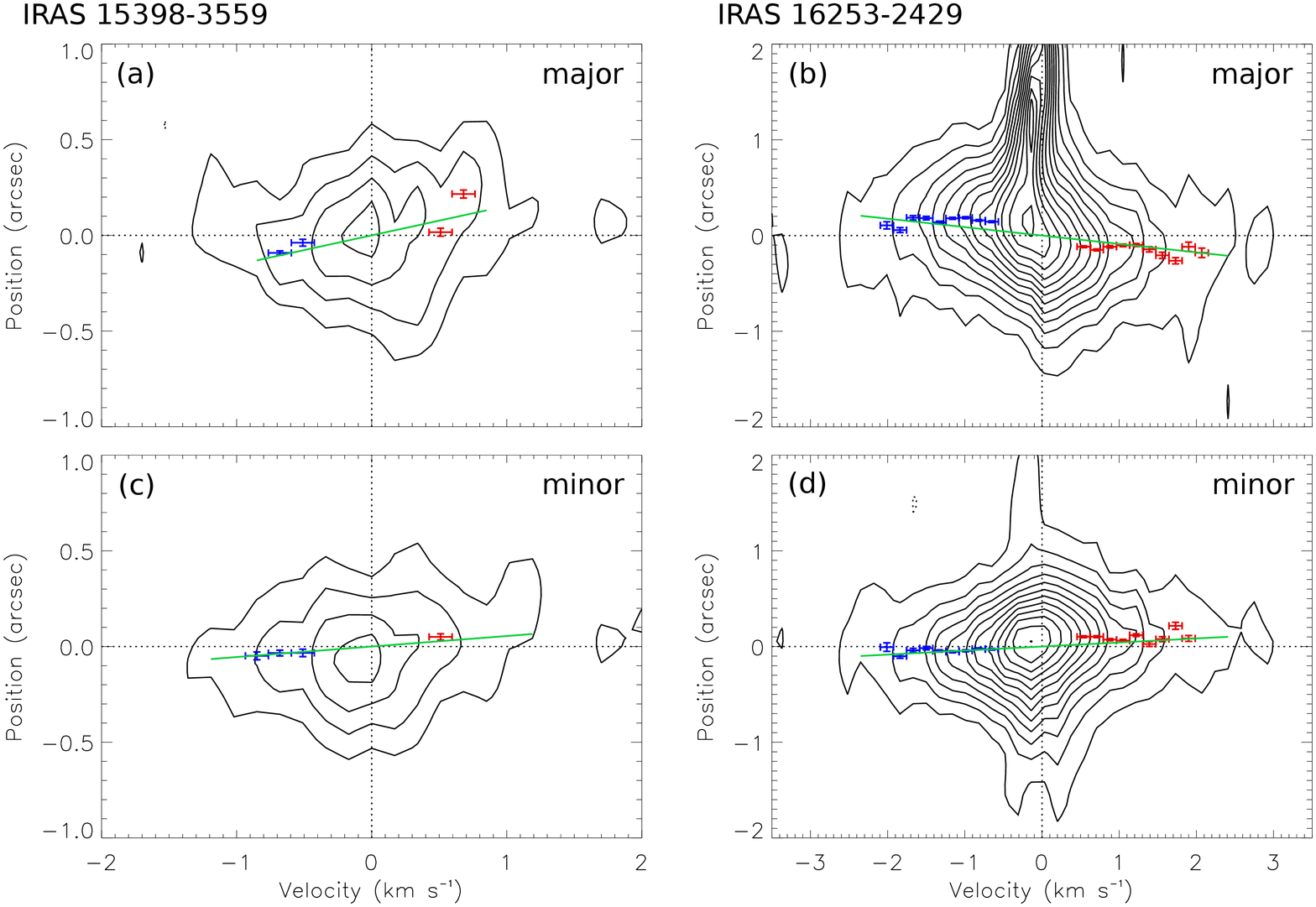}
\caption{P--V diagrams of the SO emission in IRAS 15398$-$3559 (a \& c) and in IRAS 16253$-$2429 (b \& d) observed with ALMA. Top and botton rows present the P--V diagrams along the major and minor axes, respectively. Green lines present the measured velocity gradients from the blue and red data points. Contour levels all start from 3$\sigma$, and are in steps of 3$\sigma$ in (a) and (c) and 5$\sigma$ in (b) and (d), where the 1$\sigma$ levels are listed in Table \ref{obs}.}\label{sopv}
\end{figure*}

\subsubsection{IRAS 15398$-$3559}
The rotational profile in IRAS 15398$-$3559 is $V_{\rm rot} = (0.17\pm0.02) \times (R_{\rm rot}/R_0)^{-1.0\pm0.06}$ with $V_{\rm sys} = 5.24\pm0.03$ km s$^{-1}$ (Fig.~\ref{c18opv}d), where $R_0$ is 1$\arcsec$ (150 AU in IRAS 15398$-$3559). 
This estimated $V_{\rm sys}$ is slightly larger than the single-dish measurement of 5.15 km s$^{-1}$ (Vilas-Boas et al.~2000). 
In the P--V diagram, there is additional blueshifted emission in the southeast (positive offset), 
whose velocity gradient is not the same as that of the envelope rotation. 
That component is clearly seen in the moment 0 map of the high-velocity C$^{18}$O emission ($\Delta V \gtrsim 0.3$ km s$^{-1}$; Fig.~\ref{hvmom0}b).
The blueshifted emission exhibits a secondary component with its peak position coincident with that of the redshifted component. 
Besides, clear contamination from the outflow is also observed in the blueshifted emission.
Hence, the low-resolution single-dish measurement of $V_{\rm sys}$ can be biased by these blueshifted contributions 
towards a smaller value. 
The measured power-law index of the rotational profile is consistent with a conserved angular momentum ($f = -1$) within uncertainty. 
Thus, the protostellar envelope around IRAS 15398$-$3559 is infalling with a constant angular momentum (e.g., Ulrich 1976; Takahashi et al.~2016), but its Keplerian disk is not yet observed in our ALMA observations.  
Assuming an inclination angle of 70$\arcdeg$, 
the specific angular momentum of the protostellar envelope is (1.2$\pm$0.1) $\times$ 10$^{-4}$ km s$^{-1}$ pc. 

The central compact component of the SO emission also exhibits a velocity gradient perpendicular to the outflow axis, as seen in its P--V diagram (Fig.~\ref{sopv}a), where the northwestern part is blueshifted and the southeastern is redshifted. 
The direction of this velocity gradient is identical to the one in the C$^{18}$O emission. It, thus, likely also traces the rotation of the inner envelope.  
Its velocity structure is linear-like, suggesting that it is not well resolved. 
Thus, we adopt the same method as described in Appendix \ref{vr} to measure $R_{\rm rot}$ and $V_{\rm rot}$ in the P--V diagram perpendicular to the outflow axis, 
and we fit a linear velocity gradient to the data points, 
\begin{equation}\label{mvg}
V_{\rm rot} = M_{\rm vg} \times R_{\rm rot},  
\end{equation}
where $M_{\rm vg}$ is the magnitude of the velocity gradient.
Similarly, only the high-velocity ($\gtrsim$0.4 km s$^{-1}$) channels are included, where the emission appears to be compact and has less contamination from the outflow or the extended structures (Fig.~\ref{hvmom0}c).
The magnitude of the velocity gradient perpendicular to the outflow axis is measured to be 0.04$\pm$0.003 km s$^{-1}$ AU$^{-1}$. 
The mean $R_{\rm rot}$ of the data points is 0\farcs11 (17 AU). 
With the measured $M_{\rm vg}$ and an inclination angle of 70$\arcdeg$, the rotational velocity at a radius of 17 AU is 0.9$\pm$0.05 km s$^{-1}$, corresponding to a specific angular momentum of (7.3$\pm$0.4) $\times$ 10$^{-5}$ km s$^{-1}$ pc. 
This specific angular momentum of the inner envelope traced by the SO emission is comparable to that of the C$^{18}$O rotational profile.
Additionally, the SO emission also exhibits a velocity gradient along the minor axis ($M_{\rm vg} = 0.11\pm0.007$ km s$^{-1}$ AU$^{-1}$), where the southwestern part is blueshifted and the northeastern part redshifted (Fig.~\ref{sopv}c).
The direction of this velocity gradient is the same as that of the outflow, suggesting either infall or contamination from the outflow. 
These additional observed velocity structures provide further support that rotation is not yet the dominant motion on a 100 AU scale around IRAS 15398$-$3559. 

\subsubsection{IRAS 16253$-$2429}
The P--V diagrams of both the C$^{18}$O and SO emission along the major axis of the central compact component of the 1.3 mm continuum emission in IRAS 16253$-$2429 show a small velocity gradient, where the southeastern part is blueshifted and the northwestern part redshifted (Fig.~\ref{c18opv}e \& \ref{sopv}b).  
This C$^{18}$O P--V diagram is similar to that observed in B335 (Yen et al.~2015b), where the gas motion is dominated by the infall with little rotation. 
In IRAS 16253$-$2429, the velocity structures in the C$^{18}$O emission along the major axis are not well resolved. 
There is no clear change in the peak positions from $\Delta V$ of $\sim$0.5 km s$^{-1}$ to 1.5 km s$^{-1}$ in the P--V diagram, different from those in Lupus 3 MMS and IRAS 15398$-$3559.
Therefore, the rotational profile cannot be measured. 
Moreover, this velocity gradient is not linear-like, so it cannot be well described by Equation \ref{mvg}. 
Hence, the rotational velocity of the protostellar envelope in the C$^{18}$O emission can only be estimated by using kinematic models (Section \ref{model}) but not from the P--V diagram. 
On the contrary, 
the P--V diagram of the SO emission along the major axis shows a linear-like velocity gradient. 
We adopt the same method as described above to measure $R_{\rm rot}$ and $V_{\rm rot}$ in the high-velocity ($\gtrsim$0.5 km s$^{-1}$) channels in the P--V diagram to extract a velocity gradient.
The magnitude of the gradient along the major axis is measured to be 0.091$\pm$0.004 km s$^{-1}$ AU$^{-1}$. 
The mean $R_{\rm rot}$ of the data points is 0\farcs11 (14 AU). 
With this magnitude and an inclination angle of 60$\arcdeg$, the rotational velocity and specific angular momentum in the inner envelope at a radius of 14 AU around IRAS 16253$-$2429 are estimated to be 1.4$\pm$0.6 km s$^{-1}$ and (9.6$\pm$0.4) $\times$ 10$^{-5}$ km s$^{-1}$ pc, respectively.
The P--V diagram of the SO emission along the minor axis additionally shows a velocity gradient ($M_{\rm vg} = 0.19\pm0.01$ km s$^{-1}$ AU$^{-1}$), where the southwestern part is blueshifted and the northeastern part redshifted, identical to the outflow.
This suggests that the inner envelope traced by the SO emission possibly is also infalling or affected by the outflow. 

\begin{figure*}
\figurenum{8}
\plotone{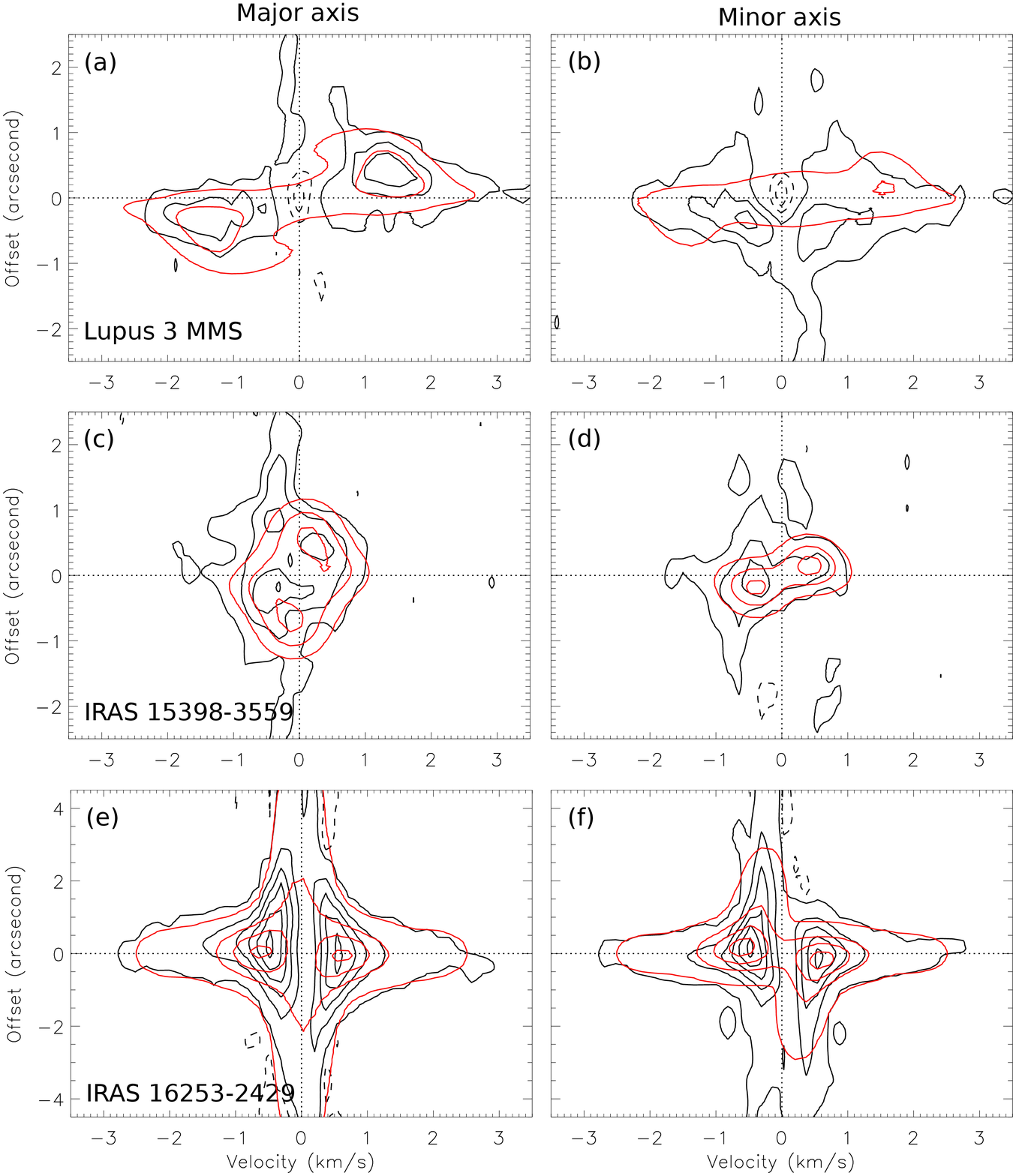}
\caption{P--V diagrams of the C$^{18}$O emission in Lupus 3 MMS (a \& b), IRAS 15398$-$3559 (c \& d), and IRAS 16253$-$2429 (e \& f) observed with ALMA (black contours) overlaid on our best-fit kinematic models (red contours). Left and right columns present the P--V diagrams along the major and minor axes, respectively. Contour levels all start from 3$\sigma$, and are in steps of 5$\sigma$ in (a)--(d) and 20$\sigma$ in (e) and (f), where the 1$\sigma$ levels are listed in Table \ref{obs}.}\label{modelpv}
\end{figure*}

\floattable
\begin{deluxetable}{lccccccccc}
\tablenum{4}
\label{fitting}
\tablewidth{0pt}
\tablecaption{Parameters of Best-fit Kinematic Models for C$^{18}$O Emission}
\tablehead{Source & PA & $i$ & $M_*$ & $R_{\rm d}$ & $j$ & $n_0$ & $T_0$ & $p$ \\
 & & & ($M_\sun$) & (AU) & (km s$^{-1}$ pc) & (cm$^{-3}$) & (K)}
\startdata
\multicolumn{9}{c}{ALMA} \\
\hline
Lupus 3 MMS & 150\arcdeg & 60\arcdeg & 0.3$^{+0.09}_{-0.05}$ & 130$\pm$70 & (9$\pm$2) $\times$ 10$^{-4}$ & (2$^{+1.5}_{-1}$) $\times$ 10$^7$ & 15$^{+40}_{-5}$ & $-2$ \\
IRAS 15398$-$3559 & 145\arcdeg & 70\arcdeg & $\leq$0.01$^{+0.02}_{-0}$ & 20$^{+50}_{-20}$ & (7$\pm$7) $\times$ 10$^{-5}$ & (2.6$\pm$0.9) $\times$ 10$^7$ & $\geq$100 & $-1.5$ \\
IRAS 16253$-$2429 & 112\arcdeg & 60\arcdeg & 0.03$\pm$0.01 & 6$\pm$6 & (6$^{+2}_{-3}$) $\times$ 10$^{-5}$ & (1.7$^{+0.7}_{0.5}$) $\times$ 10$^7$ & 55$\pm$20 & $-2.5$ \\
\hline
 \multicolumn{9}{c}{SMA} \\
\hline
 Lupus 3 MMS & 150\arcdeg & 60\arcdeg & 0.05$^{+0.2}_{-0.03}$ & $<$6 & $<$5 $\times$ 10$^{-5}$ & (1.1$^{+0.7}_{-0.4}$) $\times$ 10$^7$ & 25$^{+45}_{-20}$ & $-2$ \\
 (counter rotation) & 150\arcdeg & 60\arcdeg & $\leq$0.01$^{+0.14}_{-0}$ & \nodata & (9$^{+14}_{-7}$) $\times$ 10$^{-5}$ & (2.2$^{1.5}_{-0.7}$ $\times$ 10$^7$ & 75$^{+50}_{-45}$ & $-2$ \\
IRAS 15398$-$3559 & 145\arcdeg & 70\arcdeg & 0.02$^{+0.07}_{-0.01}$ & $<$1200 & (1$^{+4}_{-1}$) $\times$ 10$^{-4}$ & (1.4$^{+1.7}_{-0.4}$) $\times$ 10$^7$ & 15$^{+65}_{-10}$ & $-1.5$ \\
IRAS 16253$-$2429 & 112\arcdeg & 60\arcdeg & 0.02$^{+0.02}_{-0.01}$ & $<$970 & (2.3$^{+2.2}_{-2.3}$) $\times$ 10$^{-4}$ & (4.5$^{+2.7}_{-1.6}$) $\times$ 10$^7$ & 15$\pm$10 & $-2.5$ \\
\enddata
\tablecomments{PA, $i$, and $p$ are fixed parameters. The finest steps in parameters to compute models are 0.01 $M_\sun$ for $M_*$, 1 $\times$ 10$^{-5}$ km s$^{-1}$ pc for $j$, 1 $\times$ 10$^6$ cm$^{-3}$ for $n_0$, and 5 K for $T_0$. We only search for best-fit models with $T_0$ up to 100 K. $R_{\rm d}$ in these three protostars cannot be measured with our SMA data. For the fitting results of the SMA C$^{18}$O data, we only present the upper limits of $R_{\rm d}$ computed with the upper bounds of $j$ and the lower bounds of $M_*$.}
\end{deluxetable}

\subsection{Kinematic Models of Envelopes}\label{model}
To further constrain envelope rotation, protostellar mass, and disk size in these three protostars, 
we constructed kinematic models and computed model images in the C$^{18}$O emission. 
We then compared P--V diagrams from the model images with the observed ones. 
We adopt similar kinematic models and the same method 
as in Yen et al.~(2015b). 
The configuration of these models is described in Appendix \ref{kmodel}.

Figure \ref{modelpv} presents our fitting results. 
The best-fit parameters are listed in Table \ref{fitting}. 
Our fitting shows that the P--V diagram in the C$^{18}$O emission along the major axis in the Lupus 3 MMS can be explained with a Keplerian disk around a 0.3 $M_\sun$ protostar, 
and the disk radius is derived to be 130 AU (Fig.~\ref{modelpv}a).
That is consistent with the analytical analysis of the P--V diagram described in Section \ref{Rot}, 
although the protostellar mass derived from the fitting is 25\% larger. 
In our kinematic model for Lupus 3 MMS, the Keplerian disk is surrounded by an infalling and rotating envelope. The model P--V diagram along the minor axis shows, thus, a velocity gradient due to the infalling motion outside of the Keplerian disk (Fig.~\ref{modelpv}b). 
The observed P--V diagram along the minor axis is more complex than that from our model, which is possibly due to  outflow contamination. 
Besides, no clear velocity gradient along the minor axis is seen. 
All together,  this suggests that the gas motions in the surrounding envelope can be different from what we assumed in our kinematic models. 
Indeed, our SMA observations at an angular resolution of $\sim$7$\arcsec$ did neither detect free-fall motion toward the 0.3 $M_\sun$ protostar nor the same amount of specific angular momentum in the protostellar envelope as what is observed in our ALMA observations (Yen et al.~2015a), as will be discussed below. 
Nevertheless, no clear velocity gradient along the minor axis is an observational signature for dominant rotation. 
This is consistent with the expectation from a Keplerian disk. 

The observed velocity gradients along the major and minor axes in IRAS 15398$-$3559 and IRAS 16253$-$2429 can be well explained with our kinematic models (Fig.~\ref{modelpv}c--f). 
The derived specific angular momenta in the protostellar envelopes from our kinematic models, 7 $\times$ 10$^{-5}$ and 6 $\times$ 10$^{-5}$ km s$^{-1}$ pc, are comparable to those from the analytical analyses of the C$^{18}$O P--V diagrams and the velocity gradients of the SO emission, (7.3--12) $\times$ 10$^{-5}$ and 9.6 $\times$ 10$^{-5}$ km s$^{-1}$ pc, respectively. 
Since the Keplerian disks are not observed in IRAS 15398$-$3559 and IRAS 16253$-$2429, their protostellar masses in our kinematic models are primarily constrained by the velocity gradients along the minor axis and the line widths, on the assumption of a free-fall infalling motion. 
The protostellar masses are $\leq$0.01 $M_\sun$ and 0.03 $M_\sun$ for IRAS 15398$-$3559 and IRAS 16253$-$2429, respectively. 
ALMA observations in H$_2$CO at a similar angular resolution of $\sim$0\farcs5 also suggest a low protostellar mass of 0.02 $M_\sun$ and a disk size of less than 30 AU in IRAS 15398$-$3559 (Oya et al.~2014), consistent with our estimates. 
The precessing jet model for the CO emission in IRAS 16253$-$2429 (Hsieh et al.~2016) suggests the same central protostellar mass of 0.03 $M_\sun$ as our results.   
Our previous ALMA observations, which resolved the transition from infalling envelopes to Keplerian disks around protostars, show that the infalling velocities can be consistent with or 30\%--50\% lower than the expected free-fall velocities (Yen et al.~2014; Ohashi et al.~2014; Aso et al.~2015). 
Thus, the protostellar masses of IRAS 15398$-$3559 and IRAS 16253$-$2429 derived from our kinematic models can be considered lower limits, while the inferred radii of their Keplerian disks are upper limits. 

In IRAS 15398$-$3559, there are additional blueshifted components that have no counterparts in our kinematic models. 
These components are unlikely related to infall and rotation in IRAS 15398$-$3559. 
As it can be seen in the moment 0 map of the high-velocity C$^{18}$O emission (Fig.~\ref{hvmom0}b), 
the additional blueshifted component in the P--V diagram along the minor axis is likely associated with the outflow,    
and that along the major axis with the extended structures. 
This blueshifted component is also observed with our SMA observations and extends more than 5$\arcsec$ to the northwest (Yen et al.~2015a). 

These three protostars were also observed in the C$^{18}$O (2--1) line with the SMA at lower angular resolutions of $\sim$4$\arcsec$--7$\arcsec$, probing the gas motions on a larger scale of $\sim$1000 AU. 
With our ALMA data, the protostellar positions, the inclination angles, and the disk major axes are measured more accurately. 
Thus, we re-analyzed the SMA data with the updated parameters. 
We adopted the same kinematic models for the SMA data -- identical to the ones applied on the ALMA data -- to measure the envelope rotation on a 1000 AU scale. 
The best-fit parameters are listed in Table \ref{fitting}. 
The details of the fitting for the SMA data and the best-fit models are presented in Appendix \ref{smapv}. 

\section{Discussion}

\subsection{Keplerian Disk Formation around Lupus 3 MMS}\label{lupus3mmsdisk}
With our C$^{18}$O ALMA observations, we resolve the 100 AU Keplerian disk around Lupus 3 MMS. 
The specific angular momentum at the outer disk radius is estimated to be 9 $\times$ 10$^{-4}$ km s$^{-1}$ pc from our kinematic model. 
However, such a large amount of rotation was not observed on a 1000 AU scale with the SMA. 
We have convolved our ALMA images with the synthesized beam of our SMA observations, 
and the peak intensity in the ALMA images after convolution is below the noise level of the SMA observations. 
Therefore, we expect that our SMA observations cannot detect any Keplerian disk signature, 
and primarily trace the surrounding envelope on a 1000 AU scale.
The SMA observations show a velocity gradient along the major axis with a direction opposite to that observed on a 100 AU scale with ALMA (Fig.~\ref{smamodel}a).
As a result, the derived specific angular momentum from our kinematic model with the SMA data is orders of magnitude lower than that from the ALMA data, if the same direction of rotation is adopted in our kinematic models. 
That amount of specific angular momentum on a 1000 AU scale is insufficient to form the 100 AU Keplerian disk around the 0.3 $M_\sun$ protostar.
Opposite velocity gradients have also been observed in L1527 and TMC-1A but on relatively larger scales of $\gtrsim$3000 AU (Ohashi et al.~1997a,b). 
This could suggest the presence of counter rotation.
Alternatively, opposite velocity gradients can also result from asymmetric structures on larger scales, 
and thus, a projected velocity gradient does not represent rotation (e.g., Tobin et al.~2011).
In addition, 
the SMA observations only detect a line width of $\sim$1 km s$^{-1}$ in the C$^{18}$O emission on a 1000 AU scale, a factor of two narrower than expected from free-fall motion toward a 0.3 $M_\sun$ protostar, and there is no clear velocity gradient along the minor axis due to the infall.  
The observed features with the SMA are different from those in our kinematic model of an infalling and rotating envelope. 
If the surrounding envelope is indeed infalling, that could suggest that the infalling velocity is slower than the expected free-fall velocity on a 1000 AU scale, or that the envelope on a 1000 AU scale is asymmetric and close to be in the plane of the sky. 
The other possibility is that the surrounding envelope is not infalling but dispersing (e.g., Arce \& Sargent 2006; Takakuwa \& Kamazaki 2011; Koyamatsu et al.~2014; Takakuwa et al.~2015). 

Although the protostellar mass and the disk size in Lupus 3 MMS are similar to those in the other Class 0 protostar, L1527, 
they exhibit distinct gas kinematics from large to small scales (Yen et al.~2013, 2015a; Ohashi et al.~2014).
In L1527, 
the envelope rotation has a radial profile $\propto R^{-1}$ from 1000 AU to inner 100 AU scales and smoothly connects to the Keplerian disk. 
Besides, the signatures of the infalling motion are observed on both 1000 AU and 100 AU scales with a mass infalling rate of $\sim$1 $\times$ 10$^{-6}$ $M_\sun$ yr$^{-1}$. 
These results suggest that the envelope material likely falls toward the center with a conserved angular momentum to form a Keplerian disk around L1527. 
On the contrary, 
our SMA and ALMA observations of Lupus 3 MMS do not show such a connection between infall and rotation from envelope to disk. 
It is not clear how the Keplerian disk forms out of the protostellar envelope around Lupus 3 MMS. 
One possibility is that the parental dense core of Lupus 3 MMS might have possessed a higher angular momentum in the inner region than in the outer region, different from the expectation of typical rigid rotating cores (e.g., Goodman et al.~1993). 
Hence, once the collapse starts inside out, the angular momentum in the inner region is already sufficient to form a 100 AU Keplerian disk. 
Since there is no clear signature of infall and rotation in the protostellar envelope on a 1000 AU scale around Lupus 3 MMS, 
its Keplerian disk is unlikely to further grow by gaining more angular momentum carried inward by proceeding collapse.
Assuming all the luminosity in Lupus 3 MMS ($L_{\rm bol}$ = 0.41 $L_\sun$) is from the gravitational energy released by the accretion onto the protostar, 
its mass accretion rate ($\dot{M}_{\rm acc}$) can be estimated as
\begin{equation}\label{Macc}
\dot{M}_{\rm acc} = \frac{L_{\rm bol}R_*}{GM_*}, 
\end{equation}
where $R_*$ is the protostellar radius, adopted to be 3 $R_\sun$ (Stahler et al.~1980) and $M_*$ is the protostellar mass. 
$\dot{M}_{\rm acc}$ in Lupus 3 MMS is then 1 $\times$ 10$^{-7}$ $M_\sun$ yr$^{-1}$.
Unless Lupus 3 MMS is in a quiescent accretion phase,
at this rate it can only gain 0.1 $M_\sun$ more ($\sim$30\% of the current mass)
over the Class 0/I stage  ($<$10$^6$ yr; Enoch et al.~2009). 
It would then become a low-mass 0.4 $M_\sun$ star.

There is also a possibility that the opposite velocity gradient observed on a 1000 AU scale in Lupus 3 MMS is due to counter-rotation of the protostellar envelope.   
Theoretically, counter-rotating envelopes surrounding Keplerian disks can be caused by the Hall effect, which spins up disk-forming regions while it spins down surrounding material (Krasnopolsky et al.~2011; Li et al.~2011; Tsukamoto et al.~2015; Tsukamoto 2016). 
This can reduce the efficiency of magnetic braking and enable the formation of Keplerian disks. 
Recent MHD simulations show that counter-rotating envelopes generated by the Hall effect can have radii larger than 200 AU (Tsukamoto et al.~2015), comparable to the scale of the opposite velocity gradient in Lupus 3 MMS. 
 If we assume that the surrounding envelope observed with the SMA is indeed counter-rotating around the 100 AU disk, 
its specific angular momentum 
is 9 $\times$ 10$^{-4}$ km s$^{-1}$ pc from our kinematic models (Appendix \ref{smapv}). 
This is comparable to the specific angular momentum at the disk radius. 
Thus, the Hall effect can explain the disk formation in Lupus 3 MMS. 
Nevertheless, due to the limited angular resolution and signal-to-noise ratio of the SMA data, 
the density and velocity structures on the 1000 AU scale 
are not yet well resolved. 
No clear signature of infall and envelope rotation (along the same direction as the disk rotation) in the SMA data can also be due to asymmetric structures (if present) on a 1000 AU scale.
The projected velocity features might then not represent infall and rotation. 
Short-baseline data are needed to better unveil the gas kinematics on the 1000 AU scale around Lupus 3 MMS. 

\begin{figure}
\figurenum{9}
\plotone{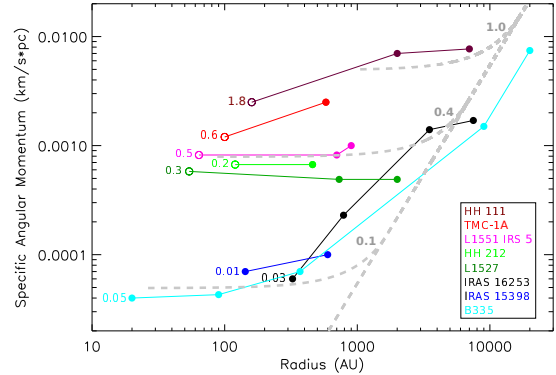}
\caption{Radial profiles of specific angular momenta. Filled and open circles display measured envelope and disk rotations in protostars from Table \ref{sample}, respectively. Each color represents a different protostar (inlet). The disk rotations are measured at outer disk radii. Grey dashed curves show the expected profiles computed from an inside-out collapse of rigid rotating dense cores where the angular momentum is conserved, for protostellar masses of 1.0, 0.4, and 0.1 $M_\sun$ from top to bottom. Numbers in colors label the measured protostellar masses of the sample protostars, as an evolutionary indicator for comparison with the expected profiles.}\label{jr}
\end{figure}

\floattable
\begin{deluxetable}{lccccccc}
\tablenum{5}
\label{sample}
\tablewidth{0pt}
\tablecaption{Comparison of Properties of Class 0 and I Protostars}
\tablehead{Source & $L_{\rm bol}$ & $\dot{M}_{\rm acc}$ & $M_*$ & $R_{\rm d}$ & $j(R)$ & $R$ & Ref. \\
& ($L_\sun$) & ($M_\sun$ yr$^{-1}$) & ($M_\sun$) & (AU) & (km s$^{-1}$ pc) & (AU)}
\startdata
HH 111 & 17.4 & 9.7 $\times$ 10$^{-7}$ & 1.8 & 160 & 2.3 $\times$ 10$^{-3}$ & 160 & 1,2\\
 & & & & & 7.0 $\times$ 10$^{-3}$ & 2000 & 2\\
 & & & & & 7.7 $\times$ 10$^{-3}$ & 7000 & 3\\
TMC-1A & 2.7 & 4.4 $\times$ 10$^{-7}$ & 0.64 & 100 & 1.2 $\times$ 10$^{-3}$ & 100 & 4,5\\
 & & & & & 2.5 $\times$ 10$^{-3}$ & 580 & 6\\
L1551 IRS 5 & 22.1 & 4.4 $\times$ 10$^{-6}$ & 0.5 & 64 & 8.2 $\times$ 10$^{-4}$ & 64 & 4,7 \\
  & & & & & 8.2 $\times$ 10$^{-4}$ & 700 & 8\\
  & & & & & 1.0 $\times$ 10$^{-3}$ & 900 & 9\\
HH 212 & 14 & 7.8 $\times$ 10$^{-6}$ & 0.2 & 120 & 6.7 $\times$ 10$^{-4}$ & 120 & 1,10\\
  & & & & & 6.7 $\times$ 10$^{-4}$ & 460 & 11\\ 
L1527 & 1.7 & 5.7 $\times$ 10$^{-7}$ & 0.3 & 54 & 5.8 $\times$ 10$^{-4}$ & 54 & 4,12\\
  & & & & & 4.9 $\times$ 10$^{-4}$ & 730 & 13,14\\ 
  & & & & & 4.9 $\times$ 10$^{-4}$ & 2000 & 15\\
IRAS 15398$-$3559 & 1.2 & 1.2 $\times$ 10$^{-5}$ & 0.01 & 20 & 7 $\times$ 10$^{-5}$ & 140 & 1,this work\\
  & & & & & 1.0 $\times$ 10$^{-4}$ & 600 & this work\\ 
IRAS 16253$-$2429 & 0.24 & 8.0 $\times$ 10$^{-7}$ & 0.03 & 6 & 6 $\times$ 10$^{-5}$ & 330 & 16,this work\\
  & & & & & 2.3 $\times$ 10$^{-4}$ & 790 & this work\\ 
  & & & & & 1.4 $\times$ 10$^{-3}$ & 3500 & 17\\ 
  & & & & & 1.7 $\times$ 10$^{-3}$ & 7500 & 17\\ 
B335 & 1.4 & 2.7 $\times$ 10$^{-6}$ & 0.05 & 3 & 4 $\times$ 10$^{-5}$ & 20 & 4,18\\
  & & & & & 4.3 $\times$ 10$^{-5}$ & 90 & 18\\ 
  & & & & & $<$7 $\times$ 10$^{-5}$ & 370 & 19\\ 
  & & & & & 1.5 $\times$ 10$^{-3}$ & 9000 & 20\\ 
  & & & & & 7.4 $\times$ 10$^{-3}$ & 20,000 & 21,22\\
\hline
Elias 29 & 14.1 & 5.7 $\times$ 10$^{-7}$ & 2.5 & 200 & 3.2 $\times$ 10$^{-3}$ & 200 & 16,23\\
R CrA IRS 7B & 4.6 & 2.0 $\times$ 10$^{-7}$ & 2.3 & 50 & 1.6 $\times$ 10$^{-3}$ & 50 & 24\\
IRS 43 & 6.0 & 3.2 $\times$ 10$^{-7}$ & 1.9 & 700 & 5.3 $\times$ 10$^{-3}$ & 700 & 16,25\\
L1489 IRS & 3.7 & 2.3 $\times$ 10$^{-7}$ & 1.6 & 700 & 2.5 $\times$ 10$^{-3}$ & 700 & 4,26\\
L1551 NE & 4.2 & 5.3 $\times$ 10$^{-7}$ & 0.8 & 300 & 2.2 $\times$ 10$^{-3}$ & 300 & 4,27\\
IRS 63 & 1.0 & 1.3 $\times$ 10$^{-7}$ & 0.8 & 170 & 1.7 $\times$ 10$^{-3}$ & 170 & 16,25\\
TMC 1 & 0.9 & 1.7 $\times$ 10$^{-7}$ & 0.54 & 100 & 1.1 $\times$ 10$^{-3}$ & 100 & 4,28\\
Lupus 3 MMS & 0.41 & 1.4 $\times$ 10$^{-7}$ & 0.3 & 130 & 9.0 $\times$ 10$^{-5}$ & 130 & 16,this work\\
L1455 IRS 1 & 3.6 & 1.3 $\times$ 10$^{-6}$ & 0.28 & 200 & 1.1 $\times$ 10$^{-3}$ & 200 & 16,28\\
VLA 1623 & 1.1 & 5,5 $\times$ 10$^{-7}$ & 0.2 & 150 & 7.9 $\times$ 10$^{-4}$ & 150 & 29\\
\enddata
\tablerefs{(1) Froebrich 2005; (2) Lee et al.~2016; (3) Lee et al.~2010; (4) Green et al.~2013; (5) Aso et al.~2015; (6) Ohashi et al.~1997a; (7) Chou et al.~2014; (8) Momose et al.~1998; (9) Saito et al.~1996; (10) Lee et al.~2014; (11) Lee et al.~2006; (12) Ohashi et al.~2014; (13) Yen et al.~2013; (14) Yen et al.~2015a; (15) Ohashi et al.~1997b; (16) Dunham et al.~2013; (17) Tobin et al.~2011; (18) Yen et al.~2015b; (19) Yen et al.~2010; (20) Yen et al.~2011; (21) Saito et al.~2000; (22) Kurono et al.~2013; (23) Lommen et al.~2008; (24) Lindberg et al.~2014; (25) Brinch \& J{\o}rgensen 2013; (26) Yen et al.~2014; (27) Takakuwa et al.~2012; (28) Harsono et al.~2014; (29) Murillo et al.~2013.}
\end{deluxetable}

\subsection{Angular Momentum Transfer From Large to Small Scales}
Our ALMA C$^{18}$O and SO observations did not detect Keplerian disks around IRAS 15398$-$3559 and 
IRAS 16253$-$2429, 
but most likely trace their infalling and rotating protostellar envelopes on a 100 AU scale. 
With our kinematic models, we measure the specific angular momenta of their rotating envelopes on a 100 AU scale 
to be 7 $\times$  10$^{-5}$ (IRAS 15398$-$3559) and 6 $\times$ 10$^{-5}$ km s$^{-1}$ pc (IRAS 16253$-$2429).
From the SMA data, the specific angular momenta of their envelopes on a 1000 AU scale are 1 $\times$ 10$^{-4}$ and 2.3 $\times$ 10$^{-4}$ km s$^{-1}$ pc, respectively. 
IRAS 16253$-$2429 has also been observed with the IRAM 30m telescope in N$_2$D$^+$, 
and exhibits a large-scale velocity gradient ($M_{\rm vg} = 1.1$ km s$^{-1}$ pc$^{-1}$) over about 15,000 AU perpendicular to the outflow axis (Tobin et al.~2011). 
The CARMA observations in N$_2$D$^+$ toward IRAS 16253$-$2429 also reveal a large-scale velocity gradient ($M_{\rm vg} = 4.1$ km s$^{-1}$ pc$^{-1}$) over about 7500 AU perpendicular to the outflow axis (Tobin et al.~2011).
These large-scale velocity gradients were interpreted as infall along the large-scale filamentary structures in IRAS 16253$-$2429 (Tobin et al.~2012b). 
If these velocity gradients are due to large-scale rotation but not infall, 
their magnitudes correspond to specific angular momenta of 1.4 $\times$ 10$^{-3}$ km s$^{-1}$ pc at a radius of 3750 AU and 1.7 $\times$ 10$^{-3}$ km s$^{-1}$ pc at a radius of 7500 AU with an inclination angle of 60$\arcdeg$.
With all these measurements, we plot specific angular momenta as a function of radius in Figure \ref{jr}.
Besides IRAS 15398$-$3559 and IRAS 16253$-$2429, we compile additional results from the literature for Class 0 and I protostars whose envelope rotations have been observed on multiple scales (Table \ref{sample}). 

In IRAS 16253$-$2429, the specific angular momentum decreases rapidly from a few thousand AU to a few hundred AU. 
This profile is similar to the one observed on the same scales in B335. 
These steep profiles from large to small scales are likely associated with the initial distributions of angular momenta in the parental dense cores (e.g., Takahashi et al.~2016). 
No inner flat profile, as observed within a few hundred AU in B335, is seen in IRAS 16253$-$2429. 
This suggests that our observations likely have not yet resolved the inner fast infalling region that causes a prolongation of infalling material and therefore, results in a flat angular momentum profile (Takahashi et al.~2016). 
In IRAS 15398$-$3559, the specific angular momentum on a 100 AU scale observed with ALMA is comparable to (or possibly smaller than) that on a scale of several hundred AU observed with the SMA, resulting in a flatter profile. 
This is also similar to the profile on approximately the same scale in B335. 
These three Class 0 protostars, IRAS 15398$-$3559, IRAS 16253$-$2429, and B335 all have low specific angular momenta ($<$10$^{-4}$ km s$^{-1}$ pc) on scales of a few hundred AU and low inferred protostellar masses$\footnotemark[3]$ ($<$0.1 $M_\sun$).
Having a low protostellar mass together with a low specific angular momentum is consistent with the expectation from an inside-out collapse of a rigid rotating dense core. 
In Figure \ref{jr}, we compare the observed profiles with the expected profiles from the inside-out collapse model (grey dashed curves) which are computed following Yen et al.~(2011, 2013).
A mean observed angular velocity of dense cores around protostars of 7.5 $\times$ 10$^{-14}$ s$^{-1}$ (e.g., Tobin et al.~2011) and a sound speed of 0.23 km s$^{-1}$ are adopted in these model calculations.
We note that in this model, the angular momentum profiles on larger scales and the evolution of the profiles are closely related to the initial distributions of angular momenta in the dense cores. 
With the above typical angular velocity for core rotation and the sound speed, the specific angular momentum on a scale of a few hundred AU in the dense cores of the inside-out collapse model is expected to be 5 $\times$ 10$^{-5}$ km s$^{-1}$ pc, when the mass of the central star+disk system reaches 0.1 $M_\sun$. This is comparable to what we observe in IRAS 15398$-$3559, IRAS 16253$-$2429, and B335. 
If the initial angular velocity is higher or the sound speed is lower, the expected specific angular momentum will become larger, or it will reach the same value but with a lower star+disk mass.
In conclusion, the low protostellar masses and the low specific angular momenta of the envelope rotation on a 100 AU scale in IRAS 15398$-$3559 and IRAS 16253$-$2429 can be explained if these sources are at an early evolutionary stage. 

\footnotetext[3]{As discussed in Section \ref{model}, the protostellar masses estimated from our kinematic models are based on the assumption that the infalling motions in these protostars are free-fall. If the infalling motions are actually slower than free-fall, these masses can be a factor of a few higher.}

In addition, Figure \ref{jr} shows that those protostars having masses of $\sim$0.2--0.5 $M_\sun$ (L1551 IRS 5, HH 212, and L1527) -- which are most likely more evolved than IRAS 15398$-$3559, IRAS 16253$-$2429, and B335 -- exhibit an order of magnitude higher specific angular momenta around  (5--8) $\times$ 10$^{-4}$ km s$^{-1}$ pc and show flat angular momentum profiles on scales between 100 AU to 1000 AU (Table \ref{sample}). 
The protostellar masses and the specific angular momenta within a 1000 AU scale in these more evolved protostars are also approximately consistent with the expectation from the inside-out collapse model when the central star+disk system reaches 0.4 $M_\sun$.
Besides, the regions exhibiting roughly constant angular momenta in these more evolved protostars are larger than those in 
IRAS 15398$-$3559, IRAS 16253$-$2429, and B335. 
The similarity between profiles from observations and the inside-out collapse model supports the scenario where the fast infalling region expands as the expansion wave of collapse propagates outward,
and that more angular momentum is transferred to the inner 100--1000 AU region with the proceeding collapse. 
Furthermore, the presence of the flat angular momentum profiles in these sample sources and the smooth connection from the envelope to the disk rotation seen in L1551 IRS 5, HH 212, and L1527 could suggest that there is no efficient magnetic braking on scales within several hundred AU.
Theoretical calculations show that magnetic braking is most efficient in the region of ambipolar diffusion shocks, where the magnetic field accumulates and ions and neutrals decouple (e.g., Li et al.~2011). 
If efficient magnetic braking indeed occurs in these protostars (L1551 IRS 5, HH 212, and L1527), the location of ambipolar diffusion shocks is most likely at a radius larger than several hundred AU. 

On the other hand, 
the most evolved protostar in the present sample, HH 111, shows an observational hint of efficient magnetic braking on a 1000 AU scale. 
The ALMA and SMA observations reveal that the specific angular momentum at a radius larger than 1000 AU is a factor of three larger than that at a radius of 100 AU, where a resolved Keplerian disk is detected (Lee et al.~2016). 
Besides, HH 111 exhibits a flat angular momentum profile beyond 1000 AU, and the measured specific angular momentum is consistent with the expectation from the inside-out collapse model for a protostar larger than 1 $M_\sun$. 
A similar hint is also found in TMC-1A, where the specific angular momentum at a radius of 600 AU, measured with the NMA, is a factor of two larger than that at a radius of 100 AU from ALMA (Saito et al.~1996; Aso et al.~2014). 
Such a decrease in the specific angular momentum inside the region exhibiting a flat angular momentum profile is not observed in the other younger protostars in this sample.
However, since the number of protostars in this sample is limited, it is not yet fully clear whether the difference between HH 111 and the younger protostars is due to evolution or different physical conditions in HH 111. 

\begin{figure}
\figurenum{10}
\plotone{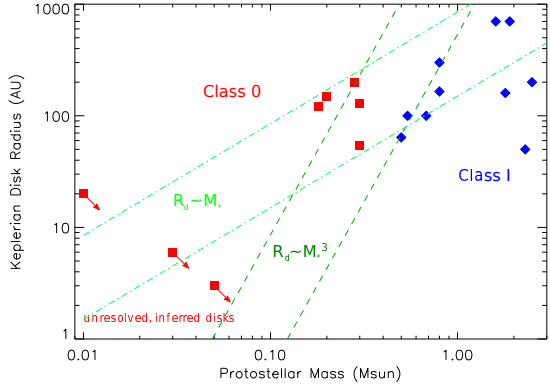}
\caption{Protostellar mass versus disk radius for the sample of protostars in Table \ref{sample}. Blue diamonds and red squares present the Class I and 0 protostars. Red squares with an arrow display the Class 0 protostars without directly observed Keplerian disks. Their inferred protostellar masses are lower limits. Their disk radii are upper limits. Dark and light green lines denote the scaling relations between protostellar mass and disk radius in the collapse models in Terebey et al.~(1984) and Basu (1998), respectively. Terebey's relation depends on the angular velocity and sound speed of dense cores. Upper and lower dark green dashed lines correspond to the relations computed with twice higher and lower angular velocities of the observed mean value (7.5 $\times$ 10$^{-14}$ s$^{-1}$; Tobin et al.~2011) and with the typical sound speed of 0.2 km s$^{-1}$. Basu's relation depends on protostar-to-disk mass ratios and the initial angular velocity and magnetic field strength in parental molecular clouds where dense cores form. Since these parameters are difficult to observationally constrain, we simply plot an upper line $850 \times (M_*/1\ M_\sun)$ AU and a lower line $150 \times (M_*/1\ M_\sun)$ AU for comparison (light green dot-dashed lines).}\label{Rd}
\end{figure}

\subsection{Evolution of Disk Size}\label{evo}
To investigate the evolution of size of Keplerian disks around protostars, we compile a list of resolved Keplerian disks around Class 0 and I protostars (Table \ref{sample}), including our ALMA results of Lupus 3 MMS, and plot their disk radii and protostellar masses measured from Keplerian rotation in Figure \ref{Rd}.
We also compare these measurements with our ALMA results of IRAS 15398$-$3559, IRAS 16253$-$2429, and B335 (Figure \ref{Rd}). 
Although the Keplerian disks around these three protostars were not resolved with our ALMA observations, 
based on the outflow/jet launching models (e.g., Shu et al.~2000; Shang et al.~2006; Machida \& Hosokawa 2013; Machida 2014), they most likely possess at least AU-scale Keplerian disks or precursors of disks (e.g., first cores)  because they are all associated with outflows and/or jets.
We have estimated their protostellar masses and inferred their disk radii based on the infall and rotation of their protostellar envelopes. 
These estimates provide a constraint on radii of Keplerian disks at an early evolutionary stage when the protostellar mass is less than 0.1 $M_\sun$.

Figure \ref{Rd} shows that the radii of the resolved Keplerian disks around Class I and 0 protostars are comparable (blue diamonds and red squares). 
Nevertheless, large disks, such as those around the Class I protostars IRS 43 and L1489 IRS, have not yet been seen around Class 0 protostars. 
Hence, these results possibly hint that disks continue to grow in size when $M_*  \gtrsim 0.2$ $M_\sun$.
To investigate the evolutionary trend of the disk sizes, 
we fitted a power-law function to the data points of resolved Keplerian disks. 
The typical uncertainties in protostellar masses estimated from Keplerian rotation are 20\%--30\% (e.g., Lommen et al.~2008; Tobin et al.~2012a, Yen et al.~2014; Aso et al.~2015), while the uncertainties in estimated disk radii are related to the relative size between the Keplerian disks and the angular resolutions (Aso et al.~2015). 
In this fitting, we adopt 30\% uncertainties in both the protostellar masses and the disk radii for all the data points, 
and we obtain
\begin{equation}\label{Rd01}
R_{\rm d} = (161\pm16) \times (\frac{M_*}{1\ M_\sun})^{0.24\pm0.12}\ {\rm AU},  
\end{equation}
where $R_{\rm d}$ is the disk radius. 
In addition, the range of disk radii around Class 0 and I protostars -- from tens to hundreds of AU -- is also comparable to that around T Tauri or Herbig Ae/Be stars (e.g., Simon et al.~2000; Pi{\'e}tu et al.~2007). 
These results are different from the expectation from the collapse models, where $R_{\rm d} \propto M_*$ or $\propto {M_*}^3$ (assuming $M_* \gg M_{\rm d}$; Terebey et al.~1984; Basu 1998).
Therefore, disk radii seem to more gradually increase with a growing protostellar mass, but not as rapidly as expected in the collapse model, and they possibly continue to grow through a transition from the Class 0 to I stage.
There are two possible scenarios to explain the slow disk growth after $M_* \gtrsim 0.2$ $M_\sun$. 
One is that the angular momentum in the outer regions ($>$ a few thousand AU) of their parental dense cores is lower than the expectation in the collapse models, where $j \propto R$ or $\propto R^2$ (Terebey et al.~1984; Basu 1998).
Thus, when the outer region starts to collapse at a later evolutionary stage, the collapsing material does not carry a large amount of angular momentum toward the center, and the disk cannot grow in size rapidly. 
Furthermore, close to the end of the main accretion phase, the mass infalling rate likely declines (e.g., Beltr{\'a}n \& de Wit 2016), and the mass reservoir is mostly already consumed or starts to dissipate, as in the case of the Class I protostar L1489 IRS (Yen et al.~2014).  
Therefore, no sufficient angular momentum is available to further grow the disk size at the later evolutionary stage. 
The second possibility is that the angular momentum transferred by the collapsing material is removed by magnetic braking. 
Hints of this removal of angular momentum are observed in HH 111. 
In this source, the specific angular momentum of the outer envelope at a radius larger than 2000 AU 
is comparable to the expectation from the collapse model (Fig.~\ref{jr}). 
However, the specific angular momentum starts to drop with decreasing radii, and the observed disk radius is only 160 AU (Lee et al.~2016). 
That disk radius is almost an order of magnitude smaller than expected assuming angular momentum conservation within 2000 AU. 
This magnetic braking effect can suppress disk growth and therefore, limit the disk size. 
Nevertheless, the signs of efficient magnetic braking, as seen in HH 111, have not been observed in the younger Class 0 and I protostars with resolved Keplerian disks, such as L1527, HH 212, and L1551 IRS 5 (Fig.~\ref{jr}). 
All together, this could suggest that magnetic braking is less of an initial condition that allows or does not allow the formation of a disk, but more of a regulatory mechanism that can control size and growth rate during the evolution of a Keplerian disk.

Our ALMA observations find two more candidate protostars exhibiting very small Keplerian disks, in addition to our previous ALMA results of B335.
In IRAS 15398$-$3559, the disk radius is estimated to be $<$20 AU, and in IRAS 16253$-$2429 $<$10 AU.
There is an order of magnitude difference in disk radii between these protostars and those exhibiting resolved Keplerian disks. 
One possibility is that there are two distinct groups of protostars that have different physical conditions, which then results in a bimodal distribution of disk radii. 
For example, MHD simulations show that the Hall effect can spin up the disk-forming region and form a large disk if the directions of the magnetic field and the rotational axis are antiparallel, while it suppresses disk formation if the two directions are parallel (e.g., Tsukamoto et al.~2015). 
These simulations demonstrate that the radii of Keplerian disks around protostars can differ by one order of magnitude because of different directions of the magnetic field, and consequently they suggest that there can be a bimodal distribution of disk radii (Tsukamoto et al.~2015).  

On the other hand, 
the inferred protostellar masses ($<$0.1 $M_\sun$) of these protostars with small disks, are all lower than those having resolved Keplerian disks with radii beyond tens of AU.
Besides, 
the protostars having small disks (IRAS 15398$-$3559, IRAS 16253$-$2429, and B335) and those having larger disks (L1527, HH 212, and L1551 IRS 5) exhibit similar angular momentum profiles. 
Their profiles can be explained with the simple collapse model (Fig.~\ref{jr}). 
Hence, the difference in their disk radii could also be an evolutionary effect, 
and disks can continue to grow in size as more angular momentum is transferred inward with proceeding collapse. 
By fitting a power-law function to the data points of only the Class 0 protostars, we obtain 
\begin{equation}\label{Rd0}
R_{\rm d} = (44\pm8) \times (\frac{M_*}{0.1\ M_\sun})^{0.8\pm0.14}\ {\rm AU}.   
\end{equation} 
Since no Keplerian rotation is detected in IRAS 15398$-$3559, IRAS 16253$-$2429, and B335, 
their protostellar masses estimated from the infalling motions are only lower limits, 
while the radii of their Keplerian disks are upper limits. 
In the Class 0 protostar L1527, the infall velocity is only 50\% of the free-fall velocity (Ohashi et al.~2014). 
If that is also the case in these three protostars, 
their protostellar masses are underestimated by a factor of four. 
With a given angular momentum, 
the radius of a Keplerian disk is inversely proportional to its central stellar mass. 
Thus, the disk radii in these three protostars can be a factor of four smaller (e.g., Yen et al.~2015a).
In this fitting, we adopt a Monte Carlo method to estimate the uncertainties in the fitting results. 
We repeated the fitting for a 1000 times. 
Each time we randomly increased the protostellar masses of IRAS 15398$-$3559, IRAS 16253$-$2429, and B335 by a factor of one to four, and decreased their disk radii by the same factor. 
Other data points of resolved Keplerian disks were randomly varied within their 30\% uncertainties. 
With the 1000 iterations, 
the probability distributions of the fitting parameters converged, 
and we adopted the means and 1$\sigma$ widths of them as the best-fit parameters and their uncertainties. 
This fitted $R_{\rm d}$--$M_*$ relation (Eq.~\ref{Rd0}) is comparable to that in the collapse model in Basu (1998) but shallower than the one in Terebey et al.~(1984). 
The deviation between the observed relation and these two collapse models could suggest that: 
(1) the angular momentum of the collapsing material in our sample protostars is not conserved but partially removed during the collapse, different from the model assumption, 
or (2) the angular momentum profiles of the dense cores in this sample are shallower than $j \propto R$ or $\propto R^2$ in the collapse models. 
Furthermore, based on our sample, the $R_{\rm d}$--$M_*$ relation derived from the Class 0 protostars (Eq.~\ref{Rd0}) is clearly steeper than that from the Class 0 and I protostars whose $M_* \gtrsim 0.2$ $M_\sun$ (Eq.~\ref{Rd01}). 
If our sample protostars are representative to probe the various evolutionary stages, 
these results likely suggest that Keplerian disks grow rapidly from $<$10 AU to tens of AU when protostellar masses increase from $<$0.1 $M_\sun$ to $\sim$0.2 $M_\sun$. 
After that, in the later evolutionary stage, rapid disk growth is suppressed.
However, the number of the protostars with $M_* < 0.1$ $M_\odot$ in our sample is only three. 
A larger sample in the low-mass regime is required to reveal the genuine $R_{\rm d}$--$M_*$ relation and the trend of disk growth.

\begin{figure*}
\figurenum{11}
\plotone{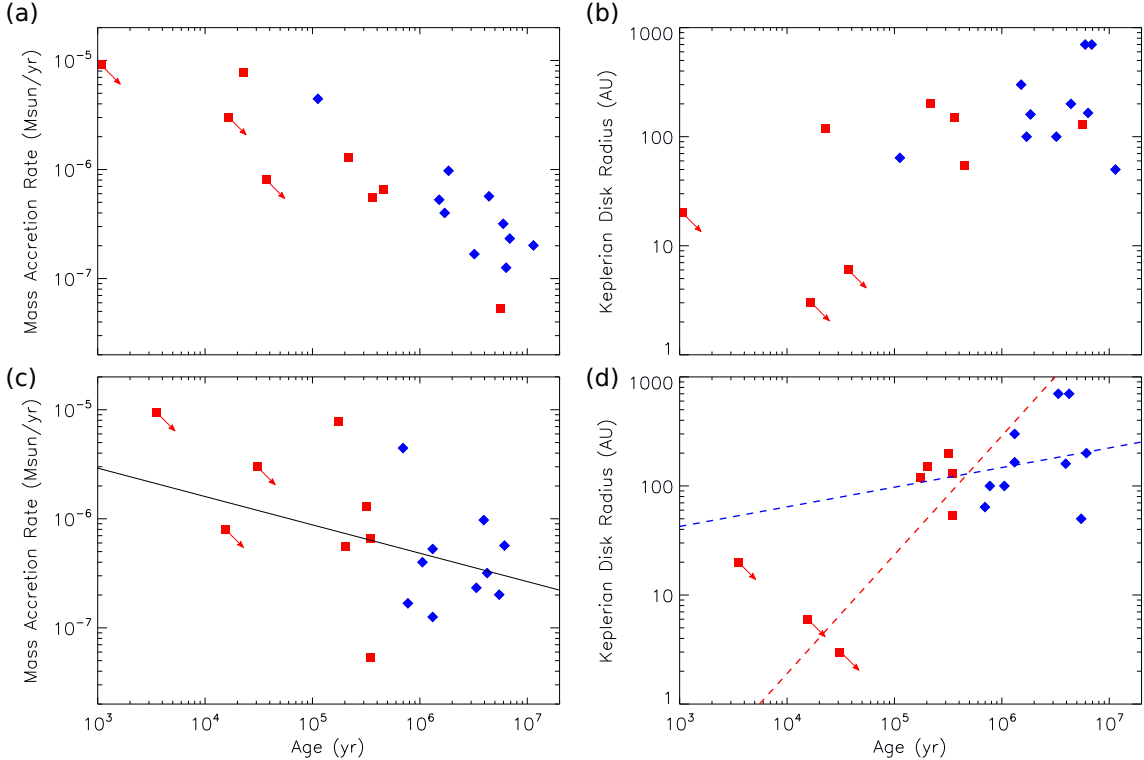}
\caption{Mass accretion rates (a \& c) and disk radii (b \& d) as a function of age for the sample of protostars in Table \ref{sample}. Blue diamonds and red squares present the Class I and 0 protostars. Red squares with an arrow display the Class 0 protostars without directly observed Keplerian disks, and their estimated mass accretion rates and Keplerian disk radii are upper limits while ages are lower limits. In (a) \& (b), the age is derived assuming that the mass accretion rate in each protostar is constant. (c) \& (d) present the results after a converged iteration, assuming the mass accretion rate declines as a power-law function of age (see Section \ref{evo}). The solid line in (c) denotes the derived mass accretion rate as a power-law function of age (Eq.~\ref{Mdot}) after convergence. Blue and red dashed lines in (d) present the fitted power-law functions to the data points of the Class 0 and I protostars exhibiting resolved Keplerian disks (Eq.~\ref{Rt01}) and the Class 0 protostars only (Eq.~\ref{Rt0}), respectively.}\label{age}
\end{figure*}

In the conventional collapse models in Terebey et al.~(1984) and Basu (1998), 
the mass accretion rate is constant, 
and the age ($t_{\rm age}$) of a protostar can be estimated as 
\begin{equation}\label{tage}
t_{\rm age} = M_* / \dot{M}_{\rm acc}.
\end{equation} 
Hence, their $R_{\rm d}$--$M_*$ relations 
imply a time-dependent disk growth as $R_{\rm d} \propto {t_{\rm age}}^3$ and $\propto t_{\rm age}$, respectively. 
In order to observationally constrain the time dependence of this disk growth, 
we first derive the age of each protostar from its protostellar mass and mass accretion rate with Equation \ref{tage} on the assumption of constant mass accretion rates, 
even though some observations have shown that the mass accretion rate declines with evolution (e.g., Beltr{\'a}n \& de Wit 2016) and that it can be episodic (e.g., Dunham et al.~2008; Enoch et al.~2009; Dunham \& Vorobyov 2012). 
The accretion rates are estimated with Equation \ref{Macc} and are listed in Table \ref{sample}. 
Figure \ref{age}a and b present the derived mass accretion rates and the disk radii as a function of the estimated age. 
There is a clear trend of larger disk radii with growing age, 
and the mass accretion rate, indeed, appears to decline with age as expected.  
In a next step, in order to include this observed decline in accretion rate in our discussion, 
we assume that all the sample sources follow the same profile of mass accretion rate as a function of age. 
In other words, we assume that the sources in our sample are representative to probe the various 
evolutionary stages.
For simplicity, this profile is assumed to be a power-law function as
\begin{equation}\label{Mt}
\dot{M}_{\rm acc}(t) = \dot{M}_{\rm acc}(t_0) \times (\frac{t}{t_0})^a.
\end{equation}
This assumption implies that the differences in protostellar masses in the sample sources are due to differences in age.  
In this picture, differences in mass accretion rates of protostars with similar masses can then be associated with short-time bursts or quiescent phases of mass accretion, 
and the time-averaged accretion rate, smoothing out bursts and quiescent phases, is described by Equation \ref{Mt}. 
With this, 
the protostellar mass can be computed by integrating Equation \ref{Mt} as
\begin{equation}\label{Ms}
M_*(t_{\rm age}) = \int^{t_{\rm age}}_0 \dot{M}_{\rm acc}(t)dt.
\end{equation}
We fit Equation \ref{Mt} to Figure \ref{age}a to obtain an initial guess of $\dot{M}_{\rm acc}(t_0)$ and $a$. 
With the derived $\dot{M}_{\rm acc}(t_0)$ and $a$, 
we re-estimated the age of each protostar using Equation \ref{Ms}
to obtain a new distribution of $\dot{M}_{\rm acc}(t_{\rm age})$ versus $t_{\rm age}$. 
Then, we again fit Equation \ref{Mt} to the new distribution to obtain an updated guess of $\dot{M}_{\rm acc}(t_0)$ and $a$, resulting again in an updated estimate of $t_{\rm age}$.  
We iterated this process until $\dot{M}_{\rm acc}(t_0)$, $a$, and $t_{\rm age}$ were converging, which was with ten iterations. 
In this procedure, 
we assumed a 30\% uncertainty in the masses of the protostars with resolved Keplerian disks, 
and a factor of four uncertainty in those without resolved Keplerian disks, as discussed above. 
In addition, we assumed that the uncertainties in the bolometric luminosities are 15\% (e.g., Dunham et al.~2013; Sadavoy et al~2014). 
This results in a 33\% uncertainty and a factor of four uncertainty in the estimated mass accretion rates of the protostars with and without resolved Keplerian disks, respectively.
Note that $R_*$ of 3 $R_\odot$ are adopted for all the sample protostars, although theoretical calculations show that more massive protostars tend to have larger radii (e.g., Palla \& Stahler 1991). 
On the other hand, the bolometric luminosities of more massive protostars (presumably more evolved) can have a larger contribution from their photosphere in addition to that from accretion (e.g., D'Antona \& Mazzitelli 1994). 
These two effects, larger stellar radii and higher photosphere luminosities, could compensate each other when deriving mass accretion rates. 
Thus, we did not include these two uncertainties in our estimates for simplicity.
Then, we adopted the same Monte Carlo method as described above to estimate the uncertainties in the fitting results.
The final distributions of $\dot{M}_{\rm acc}(t_{\rm age})$ versus $t_{\rm age}$ and $R_{\rm d}$ versus $t_{\rm age}$ are presented in Figure \ref{age}c and d, 
yielding the relation
\begin{equation}\label{Mdot}
\dot{M}_{\rm acc}(t_{\rm age}) \sim (1.6\pm0.2) \times (\frac{t}{10^4\ {\rm yr}})^{-0.26\pm0.04}\ 10^{-6} \ {\rm M_\sun\ yr^{-1}}.
\end{equation} 
Our analysis suggests that the time scale of the Class 0 stage is $\sim$4 $\times$ 10$^5$ yr, which is comparable to that estimated from the number counts of Class 0 protostars (Enoch et al.~2009). 
We also fit a power-law function to the distribution of $R_{\rm d}$ versus $t_{\rm age}$ for the Class 0 and I protostars exhibiting resolved Keplerian disks, and we obtain
\begin{equation}\label{Rt01}
R_{\rm d} = (125\pm16) \times (\frac{t_{\rm age}}{t_{\rm d}})^{0.18\pm0.09}\ {\rm AU}.
\end{equation}
For all the Class 0 protostars in the sample we find
\begin{equation}\label{Rt0}
R_{\rm d} = (106\pm46) \times (\frac{t_{\rm age}}{t_{\rm d}})^{1.09\pm0.37}\ {\rm AU}, 
\end{equation}
where $t_{\rm d}$ is 4 $\times$ 10$^5$ yr. 
In conclusion, our results suggest that 100 AU Keplerian disks likely form at the Class 0 stage within a time scale of  $\sim$4 $\times$ 10$^5$ yr, 
and then, the disk growth rate declines towards the end of the Class 0 stage. 

\section{Summary}
We perform imaging and analyses on our ALMA cycle-2 data of the 1.3 mm continuum, $^{12}$CO (2--1), C$^{18}$O (2--1), SO (5$_6$--4$_5$) emission in three Class 0 protostars, Lupus 3 MMS, IRAS 15398$-$3559, and IRAS 16253$-$2429. The aim is to probe their Keplerian disks and gas kinematics on a 100 AU scale, and to investigate formation and growth of Keplerian disks around protostars. Our main results are summarized below.
\begin{enumerate}
\item{The $^{12}$CO emission in our ALMA observations primarily traces the outflows in these protostars. By quantitatively comparing their observed morphologies and velocity structures with analytical functions of the wind-driven outflow model, the inclination angles in Lupus 3 MMS, IRAS 15398$-$3559, and IRAS 16253$-$2429 are estimated to be 60\arcdeg, 70\arcdeg, and 60\arcdeg, respectively.}

\item{A compact continuum component with a deconvolved size of 0\farcs39 $\times$ 0\farcs23 ($\sim$80 AU $\times$ 50 AU) is observed in Lupus 3 MMS, while no significant extended emission is detected on a 1000 AU scale as shown by the visibility amplitude profile. The rotational profile of the compact component traced by C$^{18}$O is measured to be $\propto R^{-0.57\pm0.03}$, consistent with Keplerian rotation. These results suggest that a 100 AU Keplerian disk has formed in Lupus 3 MMS. With our kinematic models, the protostellar mass and the outer disk radius are estimated to be 0.3 $M_\sun$ and 130 AU, respectively. SO is also detected toward Lupus 3 MMS, but appears not to be associated with the inner envelope but more likely with the outflows.}

\item{Compact ($<$30 AU) components embedded in 1000 AU extended structures are detected in the 1.3 mm continuum emission in IRAS 15398$-$3559 and IRAS 16253$-$2429. No Keplerian rotation is observed toward these two protostars. The C$^{18}$O emission on a 100 AU scale in IRAS 15398$-$3559 shows a rotational profile $\propto R^{-1\pm0.06}$, consistent with infall with constant angular momentum. On the contrary, the C$^{18}$O rotational profile in IRAS 16253$-$2429 remains unresolved in our observations. With our kinematic models for C$^{18}$O, protostellar mass, disk radius, and specific angular momentum of the envelope rotation on a 100 AU scale are estimated to be 0.01 $M_\sun$, 20 AU, and 7 $\times$ 10$^{-5}$ km s$^{-1}$ pc in IRAS 15398$-$3559, and 0.03 $M_\sun$, 6 AU, and 6 $\times$ 10$^{-5}$ km s$^{-1}$ pc in IRAS 16253$-$2429, on the assumptions of conserved angular momentum and free-fall infalling motion. SO associated with the inner envelopes is observed in both protostars, and shows signs of envelope rotation with specific angular momenta comparable to those observed in C$^{18}$O.}

\item{The protostellar envelope on a 1000 AU scale around Lupus 3 MMS in C$^{18}$O observed with the SMA shows a velocity gradient opposite to the direction of disk rotation seen with ALMA, and a line width twice narrower than the expectation from free fall toward a 0.3 $M_\sun$ protostar. Comparing our SMA and ALMA results, the narrow line width on a 1000 AU scale could suggest that the infalling velocity in the 1000 AU envelope is slower than the free-fall velocity, or that the envelope surrounding the disk is dissipating but not infalling. The opposite velocity gradient observed with the SMA could indicate that the surrounding envelope is counter-rotating with respect to the Keplerian disk seen with ALMA. If the surrounding envelope is, indeed, counter-rotating, the formation of the Keplerian disk in Lupus 3 MMS can be related to the Hall effect. Another possibility is that the protostellar envelope on a 1000 AU scale around Lupus 3 MMS is asymmetric, and the observed velocity gradient does not correspond to infall nor rotation.}

\item{Together with our SMA results and those from the literature, the radial profiles of specific angular momenta from thousands to hundreds of AU in IRAS 15398$-$3559 and IRAS 16253$-$2429 are revealed. IRAS 15398$-$3559 shows a shallow angular momentum profile on a scale of hundreds of AU, while IRAS 16253$-$2429 displays a steep profile from thousands of AU to the inner few hundred AU. These angular momentum profiles can be explained with the inside-out collapse model, if these two protostars are at an early evolutionary stage. In comparison with the angular momentum profiles of the entire sample of 8 Class 0 and I protostars, we find that the overall evolutionary trend can be described with the conventional inside-out collapse model. In addition, the angular momentum profiles in this sample could suggest that the region of efficient magnetic braking is likely located at a radius beyond several hundred AU, if magnetic braking, indeed, can efficiently remove angular momentum from infalling material.
}

\item{Our ALMA Observations of IRAS 15398$-$3559 and IRAS 16253$-$2429 and our previous observations of B335 constrain the radii of Keplerian disks around Class 0 protostars with masses of less than 0.1 $M_\sun$ to be smaller than 10--20 AU. We have compared protostellar masses, disk radii, and bolometric luminosities of these three protostars with other Class 0 and I protostars exhibiting resolved Keplerian disks. The results suggest that the size of Keplerian disks around protostars can grow more rapidly as $R_{\rm d} \propto {M_*}^{0.8\pm0.14}$ or $\propto {t_{\rm age}}^{1.09\pm0.37}$ at the Class 0 stage, where $R_{\rm d}$ is the disk radius, $M_*$ is the protostellar mass, and $t_{\rm age}$ is the age of the protostars. Consequently, 100 AU Keplerian disks likely form at the Class 0 stage within $\sim$4 $\times$ 10$^5$ yr. Then, the disk growth rate slows down as $R_{\rm d} \propto {M_*}^{0.24\pm0.12}$ or $\propto {t_{\rm age}}^{0.18\pm0.09}$ at the Class I stage. We also find an observational hint of a declining mass accretion rate $\propto {t_{\rm age}}^{-0.26\pm0.04}$ from the Class 0 to I stage. The derived disk growth rate from the observed distributions of disk sizes is lower than in the conventional collapse models where the angular momentum is conserved in Terebey et al.~(1984) and Basu et al.~(1998). This could suggest that the angular momentum of the infalling material is partially removed during the collapse, or that the initial angular momentum profiles of dense cores are shallower than those in the models.}
\end{enumerate}

\acknowledgments
This paper makes use of the following ALMA data:    ADS/JAO.ALMA\#2013.1.00879.S. ALMA is a partnership of ESO (representing    its member states), NSF (USA) and NINS (Japan), together with NRC (Canada) and NSC and ASIAA (Taiwan) and KASI (Republic of Korea), in cooperation with the Republic of Chile. The Joint ALMA Observatory is operated by ESO, AUI/NRAO and NAOJ. We thank all the ALMA staff supporting this work. 
P.M.K. acknowledges support from an Academia Sinica Career Development Award and from the Ministry of Science and Technology (MOST) of Taiwan
through grants MOST 104-2119-M-001-019-MY3. 
Y.A. is supported by the Subaru Telescope Internship Program.

\appendix

\section{Measuring Outflow Orientation and Inclination}\label{wind}
We adopt the wind-driven-shell model (e.g., Shu et al.~1991, 2000) to compare with our observed morphologies and velocity structures of the outflows and to estimate their orientations and inclinations. 
An axisymmetric model of a wind-driven outflow is described in Lee et al.~(2000) in cylindrical coordinates ($R$, $z$) as 
\begin{equation}\label{windeq}
z = c_0 R^2,\ V_R = v_0 R,\ {\rm and}\ V_z = v_0 z, 
\end{equation}
where $z$ is the distance to the protostar along the outflow axis, $R$ is the distance perpendicular to the outflow axis, and $V_z$ and $V_R$ are the velocities along the $z$ and $R$ direction. 
This model has two parameters, $c_0$ and $v_0$ in units of arcsec$^{-1}$ and km s$^{-1}$ arcsec$^{-1}$, to describe the morphology and velocity structure of an outflow, and two additional parameters, the inclination angle $i$ and the position angle $PA$, to project and rotate the outflow model in the plane of the sky.
In the present paper, $i$ is defined as the angle between the disk plane and the plane of the sky, i.e., $i = 90\arcdeg$ corresponds to the edge-on geometry with the outflow axis being in the plane of the sky. 

Below we describe our process to measure outflow orientations and inclinations. 
For simplicity, we only compare morphology and velocity structures captured by Equation \ref{windeq}, without generating model images of outflows specifically for $^{12}$CO, which would require more sophisticated models including density distributions and excitation.  
We first fixed $i$ to be 90$\arcdeg$ and searched for the best $c_0$ and $PA$ in steps of 5$\arcdeg$ to describe the outflow morphologies observed in the $^{12}$CO moment 0 maps (Fig.~\ref{lupus3mms}a, \ref{i15398}a, and \ref{i16253}a). 
To quantitatively compare the observed morphologies and Equation \ref{windeq}, 
we arbitrarily assigned a value one to pixels above 5$\sigma$ in the $^{12}$CO moment 0 maps.
We then generated simulated maps having the same size and the same pixel size as those of the observed moment 0 maps. 
For given $i$, $c_0$, and $PA$, we assigned the same arbitrary value one to pixels within the parabolic curve described by Equation \ref{windeq} in the simulated map. 
Next, we subtracted the simulated maps from the observed ones, and we computed the sum of the residuals to quantify their differences.
This allowed us to quantitatively compare the observed morphologies and Equation \ref{windeq} without involving detailed $^{12}$CO intensity distributions. 
This leads to a best-fit $PA$. 
We note that the residual gradually increases by 10\%--30\% when the PA deviates from the best fit by 5$\arcdeg$. 
The residual starts to quickly increase when the PA is more than 10$\arcdeg$ different from the best fit. 
Hence, the uncertainties in the best-fit PA are estimated to be 5$\arcdeg$.
After fixing $PA$, we searched for $c_0$ that best described the outflow morphologies for every $i$ in step of 5$\arcdeg$ with the $^{12}$CO moment 0 maps. 
With this, we obtained a series of pairs of $c_0$ and $i$. 
For each ($c_0$, $i$) pair, we then searched for $v_0$ that best described the velocity structures of the outflows observed in the $^{12}$CO P--V diagrams (Fig.~\ref{copv}). 
Similarly, we assigned pixels above 5$\sigma$ in the observed P--V diagrams to be one, generated simulated P--V maps, and again assigned pixels within the curves described by Equation \ref{windeq} for given $c_0$, $i$, and $v_0$ in the simulated maps to be one. 
Differences were then again quantified by the sum of the residuals after subtracting the simulated P--V maps from the observed ones. 
With this process, we obtained a best-fit $v_0$ for each ($c_0$, $i$) pair. Eventually, we selected the one set ($c_0$, $i$, $v_0$) that showed the minimum difference between model and observation as our final estimate (Table \ref{outflowfit}).
Since the orientations of the blue- and red-shifted outflows may not be fully aligned -- as seen in the  $^{12}$CO moment 0 maps --
we treated the blue- and red-shifted outflows separately in our fitting process.
Nevertheless, although the axes of the blue- and red-shifted outflows in these protostars are measured to be misaligned by $\sim$5$\arcdeg$--25$\arcdeg$, 
their estimated $i$, $c_0$, and $v_0$ are still comparable. 
Because of the projection effect, the velocity structures of the outflow models with different sets ($i$, $c_0$, $v_0$) can appear to be similar. 
In Table \ref{outflowfit}, 
we list the parameter ranges where our outflow models have $<$15\% larger residuals compared to the best fits as the errors of $i$, $c_0$, and $v_0$. 
Note that these errors only represent the degeneracy between $i$, $c_0$, and $v_0$ but not measurement errors.

Our results show that the wind-driven outflow model can well explain the observed outflow morphologies in the $^{12}$CO emission in Lupus 3 MMS and IRAS 16253$-$2429 and also that at the base in IRAS 15398$-$3559.  
The heads of the $^{12}$CO outflow in IRAS 15398$-$3559 show bow-like structures and cannot be explained by this wind-driven outflow model,
suggesting that this outflow 
might be composed of two components, namely a wind-driven outflow at the base and jet-driven bow shocks at the heads. 
The previous SMA observations of the outflow in IRAS 15398$-$3559 at lower angular resolutions also indicate the presence of jet-driven bow shocks in addition to a wind-driven outflow (Bjerkeli et al.~2016). 
The overall velocity structures revealed in the $^{12}$CO P--V diagrams in these three protostars, showing fan-like structures, can also be explained by the wind-driven outflow model, although the velocity structures around their systemic velocities are less clear because of missing flux. 
In the $^{12}$CO P--V diagram of IRAS 15398$-$3559, there is a component showing a roughly constant velocity along the outflow axis while exhibiting a wide velocity range at the head. This is on top of the fan-like structures at the base.  
The velocity structures of the additional component in the outflow in IRAS 15398$-$3559 is similar to that of jet-driven bow shocks (e.g., Lee et al.~2000), supporting the presence of jet-driven bow shocks in IRAS 15398$-$3559. 
This bow shock component is excluded in our process to measure outflow orientation and inclination with the wind-driven outflow model. 

\section{Measuring Rotational Profiles}\label{vr}
We adopt the method described in Yen et al.~(2013) to measure the rotational profiles in the three protostars, 
Lupus 3 MMS, IRAS 15398$-$3559, and IRAS 16253$-$2429.
Here,  we briefly summarize the process. For more details of the method, we refer to Yen et al.~(2013).
We first made P--V diagrams of the C$^{18}$O emission along the major axis of the central compact continuum components.
For IRAS 15398$-$3559, the central continuum component is not resolved, 
and thus, the P--V diagram was made perpendicularly to the mean direction of the axes of the blue- and red-shifted outflows (145$\arcdeg$, Table \ref{outflowfit}).
Then, we measured the peak positions at given velocity channels in the P--V diagrams (Fig.~\ref{c18opv}a \& c). 
The measured peak positions were adopted as $R_{\rm rot}$, and the relative velocities ($\Delta V$) to $V_{\rm sys}$ at their velocity channels ($= |V_{\rm LSR} - V_{\rm sys}|$) as $V_{\rm rot}$ (Fig.~\ref{c18opv}b \& d). 
The velocity channels close to the systemic velocities were excluded because there the velocity gradients tend to become linear, suggesting that the velocity structures are not well resolved. 
The moment 0 maps of the velocity channels at higher velocities included in this process are shown in Figure \ref{hvmom0}.
We fitted these data points with a power-law function, 
\begin{equation}
|V_{\rm LSR}(R_{\rm rot}) - V_{\rm sys}| = V_{\rm rot}(R_0) \cdot (\frac{R_{\rm rot}}{R_0})^{f}, 
\end{equation}
where $R_0$ is an arbitrary characteristic radius and is set to be 1$\arcsec$, and $f$ is the power-law index of the rotational profile. 
We included $V_{\rm sys}$ as a free parameter. 
In other words, we assumed the blue- and red-shifted C$^{18}$O emission traces the same rotational profile to measure $V_{\rm sys}$. 
Hence, there are three free parameters in the fitting, $V_{\rm rot}(R_0)$, $f$, and $V_{\rm sys}$. 
To estimate the uncertainties of our best-fit values, 
we adopt a Monte Carlo method. 
We repeated this fitting process for a 1000 times, which is sufficient to reach convergence, and each time we randomly varied the data points 
within their uncertainties. 
Then we measured the 1$\sigma$ widths of the obtained probability distributions of the best-fit parameters, and adopted them as the 1$\sigma$ uncertainties. 
Other uncertainties due to possible contamination from infalling motion and limited angular resolutions are less than 10\%--20\%, and are discussed in Yen et al.~(2013) and Aso et al.~(2015).

\section{Configuration of Kinematic Model}\label{kmodel}
The details of our model and calculations are described in Yen et al.~(2015b). 
Below, we summarize our kinematic model.
The density ($n$) and temperature ($T$) profiles in our kinematic model were adopted to be power-law functions, as
\begin{equation}\label{nr}
n(r) = n_0 \times (\frac{r}{100\ {\rm AU}})^p, 
\end{equation}
and
\begin{equation}\label{tr}
T(r) = T_0 \times (\frac{r}{100\ {\rm AU}})^q, 
\end{equation}
where $r$ is the radius. 
$q$ was fixed to be $-0.4$, the typical power-law index of  temperature profiles in protostellar sources (Shirley et al.~2000). 
We performed fittings with three different $p$ ($-1.5$, $-2$, and $-2.5$), compared the best-fit models of each $p$, and selected the one that best matched the observed velocity structures (i.e., having lowest $\chi^2$). 
The outer radius was set to be the observed radius of the central compact component in C$^{18}$O. 
This observed radius was measured by fitting a 2-dimensional Gaussian to the C$^{18}$O moment 0 maps (Fig.~\ref{lupus3mms}d, \ref{i15398}d \& \ref{i16253}d).
In Lupus 3 MMS and IRAS 16253$-$2429, the inclination angles estimated from the outflows are consistent with those inferred from the 1.3 mm continuum emission within 10$\arcdeg$ (Section \ref{outflow}). The latter ones are derived on the assumption that the continuum components are geometrically thin circular disks (Section \ref{outflow}). 
This suggests that the inner envelopes on a 100 AU scale around Lupus 3 MMS and IRAS 16253$-$2429 are flattened. 
Although the central compact component of the 1.3 mm continuum emission is not resolved in IRAS 15398$-$3559, 
the aspect ratios of the major and minor axes of the SO and the central C$^{18}$O component are larger than two, indicating that the inner envelope around IRAS 15398$-$3559 is also flattened. 
To mimic the flattened envelopes
in our model, we set $n(r)$ in the region within 80$\arcdeg$ from the polar axis to be zero, as the difference in the inclination angles estimated from the outflow and the 1.3 mm continuum emission is only $\sim$10$\arcdeg$. 
Hence, the thickness of the model envelope is linearly proportional to the radius, 
and the angle between the surface of the model envelope and the mid-plane is 10\arcdeg.
Considering the simplicity of the assumed density distribution in our model, $p$ of the best-fit model may not represent the true power-law index of the volume density profile, but more likely our model only mimics the observed column density profiles. 

In our model, the protostellar envelopes were assumed to be free falling toward the center with a constant angular momentum, described as 
\begin{equation}
V_{\rm in} (r) = \sqrt{\frac{2GM_r(r)}{r}}, 
\end{equation}
and
\begin{equation}
V_{\rm rot} (r) = \frac{j}{R_{\rm rot}}, 
\end{equation}
where $M_r(r)$ is the enclosed mass within a radius $r$, and $j$ is the specific angular momentum of the envelope. 
$M_r$ was computed with the central protostellar mass $M_*$ and the envelope mass from Equation \ref{nr}. 
The radius of the Keplerian disk $R_{\rm d}$ in our kinematic model was defined as the radius where the rotational velocity reaches the Keplerian velocity, 
\begin{equation}
R_{\rm d} = \frac{j^2}{GM_*}. 
\end{equation}
Within the disk radius, the gas motion was assumed to be solely Keplerian rotation without infalling motion, 
\begin{equation}
V_{\rm in} (r) = 0, 
\end{equation}
and
\begin{equation}\label{vk}
V_{\rm rot} (r) = \sqrt{\frac{GM_r(r)}{r}}. 
\end{equation}
Therefore, there are four free parameters in our kinematic model, $n_0$, $T_0$, $M_*$, and $j$. 
We computed the model images in C$^{18}$O on the assumption of local thermal equilibrium, 
and we then convolved the model images with the synthesized beams of our observations.
To quantitatively compare the velocity structures between  the model images and the observations, 
we generated P--V diagrams from the model images and performed $\chi^2$ fitting to the observed ones. 
We followed the same procedure to search for the best-fit models having a minimum $\chi^2$ as that in Yen et al.~(2015b).
To estimate the uncertainties in the best-fit parameters, 
we varied one parameter at a time and kept others fixed at the best-fit values to search for the models having a reduced $\chi^2$ of the minimum $+1$. 
Those model parameters are adopted as the uncertainties.
Although we fixed $q$ to be $-0.4$, the key parameters for our discussions, $M_*$ and $j$, are not sensitive to the choice of $q$. 
Adopting a different $q$ will change the best-fit $n_0$ and $T_0$ and the choice of $p$, as demonstrated by Yen et al.~(2015b).
We did not construct kinematic models for SO because the SO abundance and its excitation condition are more complex (e.g., Sakai et al.~2014; Ohashi et al.~2014)
and would require more sophisticated models and radiative transfer calculation.

\begin{figure*}
\figurenum{12}
\plotone{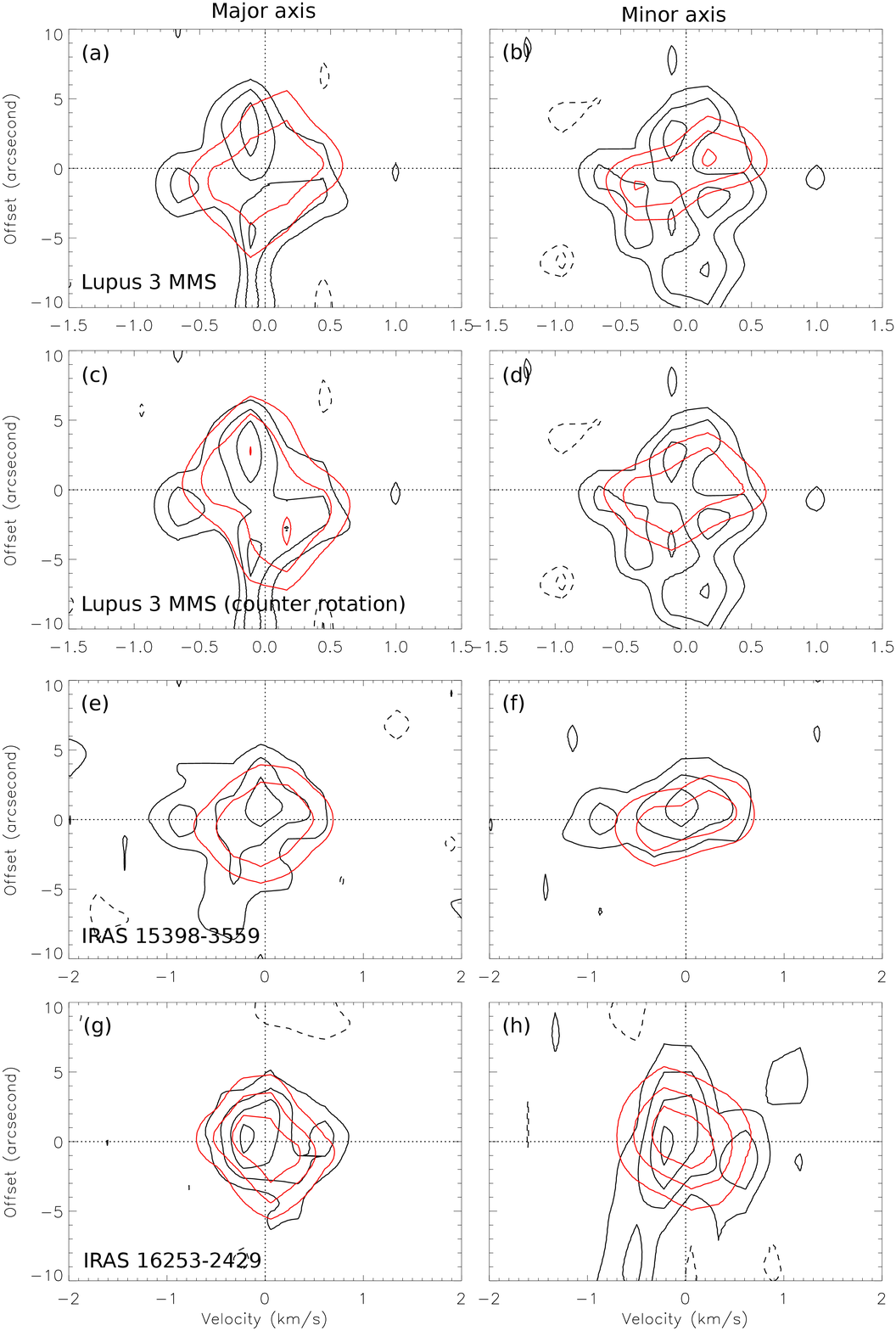}
\caption{P--V diagrams of C$^{18}$O in Lupus 3 MMS (a--d), IRAS 15398$-$3559 (e \& f), and IRAS 16253$-$2429 (g \& h) observed with the SMA (black contours), overlaid on our best-fit kinematic models (red contours). In (c) \& (d), the kinematic model has the envelope counter-rotating to the disk rotation observed with ALMA. Left and right columns present the P--V diagrams along the major and minor axes, respectively. In the diagrams of Lupus 3 MMS, contour levels start from 2$\sigma$ in steps of 1$\sigma$, where 1$\sigma$ is 0.2 Jy Beam$^{-1}$. In the diagrams of  IRAS 15398$-$3559 and IRAS 16253$-$2429, contour levels start from 2$\sigma$ in steps of 2$\sigma$, where 1$\sigma$ is 0.16 and 0.22 Jy Beam$^{-1}$, respectively.}\label{smamodel}
\end{figure*}

\section{Updated Kinematic Models for SMA Data}\label{smapv}
The SMA data are presented in detail in Yen et al.~(2015a). 
We adopted the same kinematic models and the same process described in Appendix \ref{kmodel} to fit the SMA data. 
Because of the low resolution of the SMA observations, 
the density structures of the protostellar envelopes on a 1000 AU scale around these three protostars,
Lupus 3 MMS, IRAS 15398$-$3559, and IRAS 16253$-$2429,
 were not well resolved (Yen et al.~2015a), 
and the flatness of the envelopes was not clear. 
For simplicity, we adopted the same flattened envelopes in the models, i.e., the regions within 80$\arcdeg$ from the polar axes have zero density, and identical $p$. 
Since the protostellar envelope can be counter-rotating around the Keplerian disk in Lupus 3 MMS (Section \ref{lupus3mmsdisk}), 
we also constructed models with an envelope rotation opposite to the disk rotation observed with ALMA. 
Figure \ref{smamodel} presents the P--V diagrams from our best-fit models. 
We additionally tested our fitting results against spherical envelope models.
The spherical geometry tends to smear out the velocity gradient along the minor axis induced by the infall, 
and thus, the derived protostellar mass tends to be larger and is primarily constrained by the line width. 
Nevertheless, the key parameter, the specific angular momentum, is constrained by the velocity gradient along the major axis, and is not sensitive to the geometry of the envelopes, as long as the envelopes are axisymmetric and C$^{18}$O (2--1) is optically thin. 
This test shows that the derived specific angular momentum is consistent within 10\% between the flattened and spherical envelope models.

\end{document}